\documentclass[11pt]{article}
\usepackage[dvips]{epsfig}
\newcommand{\bfmath}[1]{\mbox{\boldmath $#1$}}
\begin{document}
\title{ Kinematic simulation of multi point turbulent dispersion}
\author{M. A. I. Khan$^1$, A. Pumir$^2$ and J. C. Vassilicos$^3$\\
\small{$^1$DAMTP, University of Cambridge, Silver Street, Cambridge CB3 9EW, UK.}\\
\small{$^2$INLN, 1361 route des Lucioles, F-06560, Valbonne, France.}\\
\small{$^3$Dept. of Aeronautics, Imperial College, London SW7 2BY, UK.}}

\date{\today}
\maketitle
\begin{abstract}

As three particles are advected by a turbulent flow, they separate
from each other and develop non trivial geometries, which effectively
reflect the structure of the turbulence. We investigate here the
geometry, in a statistical sense, of three Lagrangian particles
advected, in $2$-dimensions, by Kinematic Simulation (KS). 
KS is a Lagrangian model of turbulent diffusion that
makes no use of any delta correlation in time at any level. With this
approach, situations with a very large range of inertial scales and
varying persistence of spatial flow structure can be studied.

We first show numerically that the model flow reproduces
recent experimental results at low Reynolds numbers.
The statistical properties of the shape distribution at much higher Reynolds
number is then considered. Even at the highest available inertial range,
of scale, corresponding to a ratio between large and small scales of $L/\eta
\approx 17,000$, we find that the radius of gyration of the three points does
not precisely follow Richardson's prediction. The shapes of the triangles
have a high probability to be elongated. The corresponding shape distribution
is not found to be perfectly self similar, even for our highest ratio of
inertial scales.
We also discuss how the parameters of the synthetic flow, such as the exponent
of the spectrum and the effect of the sweeping affect our results.
Our results suggest that a non trivial distribution of shapes will be observed
at high Reynolds numbers, although it may not be exactly self similar.
Special attention is given to the effects of persistence of spatial
flow structure.
\end{abstract}

\section{Introduction} 

The transport of scalar fields by turbulent flows is an important
process in many physical situations ranging from the dynamics of the
atmosphere and the ocean to chemical engineering. Specific examples of
scalars are  provided by pollutant density, temperature or humidity
fields and the concentration of chemical and biological species
\cite{shr_sig_nature2000}.

Issues of transport and  mixing in turbulence are directly related to
the properties of fluid trajectories. The problem is thus often
addressed using Lagrangian techniques \cite{yeung,Sawford2001,Pope94}.
There is an established formal connection between the statistics of
fluid particle motion and the concentration field of a diffusing
scalar \cite{GKB52}. Hence the important modeling issue of predicting
passive scalar transport in turbulence can be addressed by following
the evolution of Lagrangian particles \cite{yeung,Sawford2001,Pope94}.

The dispersion problem of one or two particles in the flow has been
studied in great detail. In particular, the seminal work of Richardson
\cite{Richardson} leads to the prediction that the separation between
two particles grows according to $\langle R ^2 \rangle \propto
\varepsilon t^3$, where $\varepsilon $ is the rate of energy
dissipation in the flow. 
Much less work has been devoted to the dispersion
of $3$ or more particles. The remarkable organization of the flow, which leads
to the formation of very sharp fronts of scalar concentration
\cite{Gibsonetal77,Mestayer82,Sreeni91,Mydlarski98}, has
a non trivial signature on the $3$-point correlation function of the flow
\cite{siggia98,Pumir98,Mydlarski98_2}. This, together with
the well-established relation between the properties of the $n$-point
correlation function and the properties of the evolution of $n$ fluid
particles advected by the flow, provides the motivation for studying
the problem of dispersion of $3$ particles or more.  Despite recent
progress both theoretically \cite{falk2001} and experimentally
\cite{Bodenschatz,Danes,Pinton,jullien99}, little is known about the
dispersion of $3$ particles or more.

An extra motivation to study dispersion of more than two particles
comes from recent theoretical attempts to model turbulent velocity
fluctuations in terms of small sets of Lagrangian particles
\cite{pumir}

The evolution of three particles configuration in turbulent flows has
been considered numerically in direct numerical simulations (DNS) of
$3$-dimensional flows, at moderate Reynolds number \cite{Pumir2000}.
Because of the limited range of inertial scales available in DNS, the
numerical studies are unable to answer questions about shape
statistics in the inertial range.  A phenomenological model,
introduced to describe the shape deformation in the inertial range, in
high Reynolds number flows, lead to the prediction of a non trivial
shape distribution \cite{Pumir2000}.  This model provided both the
motivation and the theoretical background to analyze the experimental
results of \cite{patrizia}. This experiment provided important insight
on the statistics of deformation, although the experimental setup also
suffered from the limited inertial range.

In this paper, we consider the problem of Lagrangian dispersion of $3$
particles with the help of the Kinematic Simulation (KS) method,
introduced in \cite{FHMPerk92}. KS provides a Lagrangian model of
turbulent diffusion, based on a simplified incompressible velocity
field, with a proper energy spectrum and no assumption of delta
correlation in time made at any level. This model reproduces very well
the Lagrangian properties observed in laboratory experiments
\cite{NicolVas2002}, as well as in DNS \cite{MkVas99}.  The
computational simplicity of the KS allows us to consider very large
inertial ranges : a ratio of scales of $\sim 10^4$ is easily
accessible with moderate computer resources. KS thus turns out to be
an ideal tool to study issues of dispersion in turbulent flows.

This paper is therefore devoted to the study of Lagrangian dispersion
of $3$ points in a $2$ dimensional turbulent flow using the KS. We
will demonstrate that the numerical results effectively reproduce the
experimental results of \cite{patrizia}, and we will explore the large
Reynolds number limit with the help of the KS.

In section II, we briefly discuss the parameterization used to
characterize the size and shape of the triangles, and we review the
theoretical and experimental results on shape dynamics. Technical aspects
of the simulation methods are described in section \ref{sec_ks}.  The
comparison between the experimental results of \cite{patrizia} and the KS
simulations are presented in section \ref{exp_ks_results}. Section 
\ref{large_Reynolds_limit}
contains our results concerning the large Reynolds number limit. In
section \ref{ks_new_shape} we discuss the effects of persistence
of flow structures on shape dynamics. Last, we present our concluding
remarks in section VII.

\section{Previous results on shape dynamics }

In this section, we briefly review the previous work on shape dynamics
\cite{Pumir2000,patrizia}.

\subsection{Kinematics}

The evolution of a cluster of particles is described both by the overall
scale, and by the shape of the object. In the case of a set of $n = 3$
particles, located at ${\bf x}_{i}$ ($i=1,2,3$),
we define a set of reduced vectors involving
relative separations only, defined by \cite{siggia98,Pumir98} :

\begin{eqnarray}
\label{reduced_vec1}
\bfmath{\rho}_{1}&=&{\frac{({\bf x}_{2}-{\bf x}_{1})}{\sqrt{2}}}\\
\bfmath{\rho}_{2}&=&{\frac{(2{\bf x}_{3}-{\bf x}_{2}-{\bf x}_{1})}{\sqrt{6}}}
\label{reduced_vec2}
\end{eqnarray}
The radius of gyration is defined as \cite{siggia98}
\begin{equation}
R^{2}=\sum_{i=1}^{2} \bfmath{\rho}_{i}^{2}=
{\frac{1}{3}}\sum_{jk}{\bf r}^{2}_{jk}
\end{equation}
where ${\bf r}^{2}_{ij}=|{\bf x}_{i}-{\bf x}_{j}|^{2}$ are the
distances between the vertices of the triangle. The quantity
$R$ measures the spatial extent of the swarm of particles.
In order to characterize the shape,
we introduce a moment of inertia like tensor \cite{Pumir2000}
\begin{equation}
g^{ab}=\sum_{i=1}^{2}\rho_{i}^{a}\rho_{i}^{b},
\end{equation}
where $\rho_{i}^{a}$ is the $a$-th spatial component of the vector
$\bfmath{\rho}_{i}$. For a triangle in $2$-dimensions, the tensor $g$
has two eigenvalues, $g_1 > g_2 $ (note that $g_1 + g_2 = R^2$).
These eigenvalues characterize the
spatial extent of the swarm in the two principal directions.
The ratio $I_2$ between the smallest eigenvalue, $g_2$, and $R^2$ :
\begin{equation}
I_{2} = \frac{g_{2}}{R^{2}}
\end{equation}
provides us with a quantitative measure of the shape of the object. An
equilateral triangle corresponds to $I_2 = 1/2$. The smaller $I_2$,
the more elongated the triangle is.
The moment of inertia tensor can
be used both in $2$ and $3$ dimensions to characterize a set consisting
of an arbitrary number of particles.

In the case of a triangle, a full parametrization of the shape is
provided by the quantities $w$ and $\chi$, defined by :
\begin{equation}
\chi =\frac{1}{2} \arctan \left[\frac{2 \bfmath{\rho}_1  \cdot
\bfmath{\rho}_2} {\rho_1^2-\rho_2^2} \right]\;\;\;\; ; \;\;\;\;
w=2 \,\frac{|{\bfmath{\rho}}_{1}\times{\bfmath{\rho}}_{2}|}{R^2} \label{chiw}
\end{equation}
By taking into account the symmetries of the triangle under any
reparametrization of its vertices, the parameter space is restricted
to $0 \le w \le 1$ and $ 0 \le \chi \le \pi/6$. The variables
$w$ and $I_2$ are related by the relation : $ I_{2}=(1/2)(1-\sqrt{1-w^{2}})$.
A small value of $w$ corresponds to a nearly  collinear set of points.
The quantity $\chi$ is small when the separation between
two particles, say $1$ and $2$, is much smaller than their separation
with the third one : $r_{12} << r_{13}, r_{23}$.

We consider in this work the statistical properties of the
shape distribution. To this end, we study the probability
distribution functions of the various quantities $R$,
$I_2$, $w$ and $\chi$ characterizing the shape.
The Gaussian distribution :
$P_G(\rho_1, \rho_2 ) = {\cal N} \exp \bigl( - (\rho_1^2 + \rho_2^2 ) \bigr) $
provides an interesting distribution of reference. It can be shown
\cite{patrizia} that the distributions of $\chi$ and $w$ are uniform
(in $2$ dimensions) : $P_G(\chi) = \frac{6}{\pi} $ and $P_G(w) = 1$.
In particular, the corresponding mean values are :
$\langle \chi \rangle_G = \frac{\pi}{12}$, $\langle w \rangle_G
= \frac{1}{2}$ and $\langle I_2 \rangle_G = (1 - \frac{\pi}{4})/2 $.

\subsection{Monte-Carlo model}

In order to study theoretically the distortion of sets of $3$ or $4$
particles by a turbulent flow in the inertial range of scales, a
stochastic model based on phenomenological considerations was proposed
in \cite{Pumir2000}.  At the heart of the model is a simplified scale
decomposition of the full turbulent velocity field, on the scale of
the global size of the triangle measured by the radius of gyration
$R$ \cite{Shr-Sig95}. Namely, the velocity field is written as:
\begin{equation}
{\bf v}\equiv{\bf v}_{<}+{\bf v}_{\approx}+{\bf v}_{>},
\label{vel_decomp}
\end{equation}
where ${\bf v}_{<}$ is the contribution due to the small
wavenumbers or large scales in the usual Fourier decomposition
($|k|\leq 1/2R$), ${\bf v}_{>}$ comes from the large wavenumbers
($|k|\geq 2/R$) or small scales and ${\bf v}_{\approx}$ originates
from the scales of the flow comparable to the global scale $R$
($2/R\leq |k|\leq 1/2R$).  The large scale contribution is uniform
over the triangular configuration of particles, and is therefore
assumed in \cite{Pumir2000} not to distort the set of particles.
The ${\bf v}_{\approx}$ part of the velocity field acts coherently
over the scale of the triangles with correlation time of the order
of the characteristic time of turbulence at scale $R$, defined by
:
\begin{equation}
\tau(R)=R^{2/3}\epsilon^{-1/3}
\end{equation}
The small scale component ${\bf v}_{>}$ is often assumed to be
completely incoherent on the scale $R$ of the three points and its
correlation time is short compared to the characteristic time  of
turbulence at scale $ \tau(R)$. It is modeled in \cite{Pumir2000}
by a white noise term.

The action of ${\bf v}_{\approx}$ is approximated by a
(coarse-grained) strain matrix, $M_{ab} = \partial_a v_b$, acting on
the vectors ${\bfmath \rho}_i$. The rapidly fluctuating, incoherent
component ${\bf v}_>$ is modeled by a Gaussian, white in time, random
process. This leads to the following stochastic model \cite{Pumir2000}:
\begin{eqnarray}
\frac{d\rho_{i}^{a}}{dt}&=& \rho_{i}^{b}M_{ab} + u_{i}^{a},\\
\frac{dM_{ab}}{dt}&=&-\frac{M_{ab}}{\tau(R)}+ \eta_{ab},
\label{monte_mod}
\end{eqnarray}
where the indices $i,j=1,2$ for $3$ particles labels the relative
vectors, see Eq. (\ref{reduced_vec2}), and $a,b$ labels the spatial
components.  The velocity fields ${\bf u}$ and the $\eta_{ab}$ term
are random Gaussian terms, delta-correlated in time with variances
\begin{equation}
\left\langle\eta_{ab}(t)\eta_{cd}(t')\right\rangle=C_{\eta}^{2}\delta(t-t')\left(\delta_{ac}\delta_{bd}-\frac{1}{2}\delta_{ab}\delta_{cd}\right)/\tau(R)
\end{equation}
\begin{equation}
\langle
u_{i}^{a}(t)u_{j}^{b}(t')\rangle=\left(\frac{C_{v}}{2}\right)^{2}\delta(t-t')\delta_{ij}\delta_{ab}R^{2}/\tau(R)
\label{u_variance}
\end{equation}
The stochastic model has been constructed in such a way that
the matrix $M$ is traceless (incompressibility) and correlated with a
time scale $\tau(R)$. Its amplitude is of the order of $|M|\sim R^{-1/3}$.
The dimensionless parameter $C_{v}$ (respectively $C_{\eta}$)
controls the importance of the incoherent jitter (respectively of
the coherent term) in the model.

Physically, the term $\rho_{i}^{b}M_{ab}$
in equation (\ref{monte_mod}) stretches and aligns the set of points.
This distorting action is opposed by the action of the $\bf u$ term, which
tends to make the shape distribution Gaussian. The shape distribution
resulting from these two effects is non trivial, and depends continuously
on the ratio $C_v/C_\eta$. A-priori, this number is of order $1$. In
the limit $C_v/C_\eta \rightarrow \infty$, the shape distribution becomes
Gaussian.

The model turns out to reproduce qualitatively several important aspects
of the experimental results. A detailed analysis of the experimental
data however pointed out several shortcomings of the stochastic approach
\cite{patrizia}. In particular, it was found that whereas the stochastic
model predicts a uniform distribution of the variable $\chi$, the
experimental distribution of $\chi $ shows a peak near $\chi = 0$.

\section{Kinematic simulations}\label{sec_ks}

In contrast to the stochastic model described above, KS  defines
explicitly the velocity field which advects the particles.
Following \cite{Vas98}, we define the turbulent velocity field
${\bf v}({\bf x},t)$ by summing over a set of Fourier modes, ${\bf
k}_n$ :
\begin{eqnarray}
{\bf v}({\bf x},t)=&&\sum_{n=1}^{N_{k}}\Big[{\bf A}_{n}\wedge{\bf k}_{n}\
\cos({\bf k}_{n}\cdot {\bf x}+\omega_{n}t)\nonumber\\
+&& {\bf B}_{n}\wedge{\bf k}_{n}\
\sin({\bf k}_{n}\cdot {\bf x}+\omega_{n}t)\Big]
\label{vel_KS}
\end{eqnarray}
where  $N_{k}$ is the number of modes in the simulations, ${\bf k}_{n}$
are the wave vectors, ${\bf A}_{n}$ and ${\bf B}_{n}$ are the
amplitude vectors and  $\omega_{n}$ the frequency. The norms of the
wavevector are chosen of the form
$ k_n = | {\bf k}_n | = k_0 b^n $ with a parameter $b$ typically
chosen to be \(b=(L/\eta)^{1/(N_{k}-1)}\). The large (integral)
scale, $L$, and the small (Kolmogorov) scale, $\eta$ of the flow
verify : $L = 1/k_1$ and $\eta = 1/k_{N_k}$ ($L/\eta = b^{N_k}$).
The direction of the wave vector, $\hat {k}_n = {\bf k}_n / | {\bf
k}_n | $ is uniformly distributed along the unit circle.
Similarly, the directions of the vectors ${\bf A}_n$ and ${\bf
B}_n$ are randomly distributed, and their amplitudes are chosen so
that the energy spectrum is of the form $E(k)\sim k^{-p}$. The
frequencies $\omega_n$ are taken to be
$\omega_{n}=\lambda\sqrt{k_{n}^{3}E(k_{n})}$, where $\lambda$ is a
dimensionless parameter, {\it a-priori} of ${\cal O}(1)$. The
definition Eq. (\ref{vel_KS}) makes the velocity field explicitly
incompressible. Note that no delta correlation in time is used in
KS at any level and that the parameter $\lambda$ controls the
unsteadiness of the flow.

To investigate the geometry of clusters of $n = 3$ Lagrangian
particles, we simply advect numerically Lagrangian particles in
the velocity field ${\bf v}({\bf x},t) $ defined by
Eq. (\ref{vel_KS}). This is done by solving a set of ordinary
differential equations for the position vectors ${\bf X} ({\bf x}_0, t)$ :
\begin{equation}
{\frac{d}{dt}}{\bf X}({\bf x}_{0},t)={\bf v}({\bf x}={\bf X},t)
\end{equation}
with the initial condition ${\bf X} ({\bf x}_0, 0) = {\bf x}_0$.
We start with an isotropic object, i.e. with an equilateral triangle,
of a given size, and follow its evolution over time. The quantities
characterizing the deformation of the object, such as
$I_{2}$, $w$, $\chi$ are monitored as a function of time.
We then perform ensemble averages over many triangles in different
realizations of the
velocity field to obtain the relevant particle statistics.

We firstly validate the predictions from KS with the experimental
results \cite{patrizia}. To this end, we choose the power of the
spectrum $p=5/3$, the ratio of inner to outer scales as suggested
by the experiment, and we take $\lambda = {\cal O}(1)$. We then
extrapolate our results to higher values of the ratio of inner to
outer scales $L/\eta$, to study dispersion in a high Reynolds
number flow.

We stress that KS is a model of the Eulerian velocity field, used
to advect the particles. The KS velocity field, (\ref{vel_KS}) has
an interesting spatio-temporal flow structure, which varies with
the parameters of the flow $p$ and $\lambda$ [26,27].
Investigating systematically how the changes in the parameters $p$
and $\lambda$ affects advection of particles is intrinsically
interesting in the context of this study.

\section{Validation of KS}\label{exp_ks_results}

Before making any predictions with KS regarding multi-particle
statistics we firstly validate the model by comparing with the
experimental results \cite{patrizia}. We are interested here in
reproducing {\it qualitatively} the experimental results of
\cite{patrizia}. This does not mean that KS is not able to
reproduce quantitative predictions, at the cost of fitting the
parameter $\lambda$. We are merely interested in the trends of the
shape evolution, that is, in the behaviors of the distributions of
$w$, $\chi$ etc as a function of time.

An experimental investigation of the problem of dispersion of
triangles by a turbulent flow was carried out in
$2$-dimensions, in the inverse cascade regime \cite{patrizia}. The
flow was confined in a small container, $15 \times 15 cm^2 $.
Permanent magnets were placed under the bottom of the cell. The
flow was stirred by running a current through a salted solution.
The energy was injected at the scale $l_i = 1.5 cm$. The velocity
field was recorded by using standard Particle Image Velocimetry
techniques, and was then stored on a $64 \times 64$ grid every
$0.04 s$. The resulting spatial resolution was good enough to
describe all the relevant scales of the flow. A Kolmogorov
$k^{-5/3}$ regime was observed, over the limited range of scales $
1.5 {\rm cm} \le l \le 5.5  {\rm cm} $. The time resolution was
also amply sufficient to follow numerically the evolution of
particles. The evolution of a large number ($ \sim 2 \times 10^4$)
of triangles was then followed numerically. In this section we
compare our results produced from KS with the experimental results
of \cite{patrizia} and validate our model in the process.

It was observed (see Fig.3-5 of \cite{patrizia}) that  the typical
size measured by the radius of gyration $R$ of the triangles increased
until it reached the largest scale $9cm$ of the experimental setup
where it started to fluctuate around this value. The evolution in time
of the mean values of $w$ and $\chi$ (see Figs.4,5 of
\cite{patrizia}) showed a rapid decrease of these parameters
corresponding to strong shape distortions of the triangles. The
smaller the initial separation $r_{0}$ the lower the minimal value of
this parameter was observed. The shape distortion was maximum when $R$
reached the lower value $\sqrt{R^{2}}=1.5cm$ of the inertial range.
The mean values of these variables  tend to an asymptotic value when
$R$ increases above the upper  bound $\sqrt{R^{2}}=5.5cm$ of the
inertial range. Specifically, it was found that $\langle
w\rangle_{asm}=0.5$, $\langle I_{2}\rangle_{asm}\approx 0.11$ and
$\langle\chi\rangle_{asm}\approx 0.26$ These asymptotic values for
$w$, $I_{2}$ and $\chi$ correspond to a Gaussian distribution of the
${\bfmath \rho}_{1}$ and ${\bfmath \rho}_{2}$, which implies a uniform
distribution for $w$, $I_{2}$ and $\chi$ with the Gaussian values
$\langle w\rangle_{Gau}=1/2$, $\langle
I_{2}\rangle_{Gau}=(1-\pi/4)/2=0.107$ and
$\langle\chi\rangle_{Gau}=\pi/12=0.262$ and a corresponding Gaussian
distribution for $R$: $P_{Gau}(R)=(8R^{3}/\langle
R^{2}\rangle^{2})\exp(-2R^{2}/\langle R^{2}\rangle)$.

The PDF of $R$ and $w$, see Fig.6 of \cite{patrizia}, can be well
approximated by the Gaussian distribution for large values of time
$t=80sec$ and $t=100sec$ corresponding to values of the radius of
gyration larger than the integral scale $L$. At later times, the
finite size of the experimental system induces a saturation of the
triangle size, so the tails of the distribution of $R$ could no longer
be correctly fitted by $P_{Gau}(R)$ \cite{patrizia}. A very slow
relaxation of the value of $\langle \chi \rangle$ towards its
asymptotic, Gaussian value was observed. The fact that $\chi = 0$ is
more probable than $\chi = \pi/6$ implies that triangles with one edge
much shorter than the two other ones has a large probability. This
effect should ultimately disappear at later time, in the diffusive
regime.

The numerical experiment, consists of generating KS flows in two
dimensions, with an energy spectrum $E(k)\sim k^{-5/3}$, similar to
the one observed experimentally, characterized by a ratio $L/\eta =
3.67$, and with an unsteadiness factor $\lambda = 0.5$. The smallest
and the largest time scales of the flow are defined to be
$t_{\eta}=2\pi/\sqrt{k_{\eta}^{3}E(k_{\eta})}$ and $t_{E}=L/u'$
respectively, where $u'$ is the r.m.s velocity of the flow field. In
this flow, we follow the evolution of three points, initialized as the
vertices of an equilateral triangle of size $r_0$. The results were
averaged over $\sim 10^4$ configurations.

The results obtained from KS
show the same tendencies as the one observed in the experiment,
as we now demonstrate.
\begin{figure}
\begin{center}
\includegraphics[height=7cm,width=9cm]{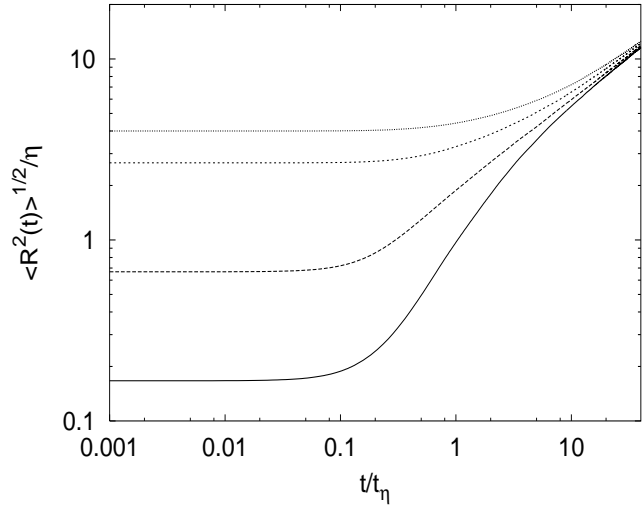}
\end{center}
\caption{Time evolution of $\langle R^{2}\rangle$}
\label{fig_1}
\end{figure}
\begin{figure}
\begin{center}
\includegraphics[height=7cm,width=9cm]{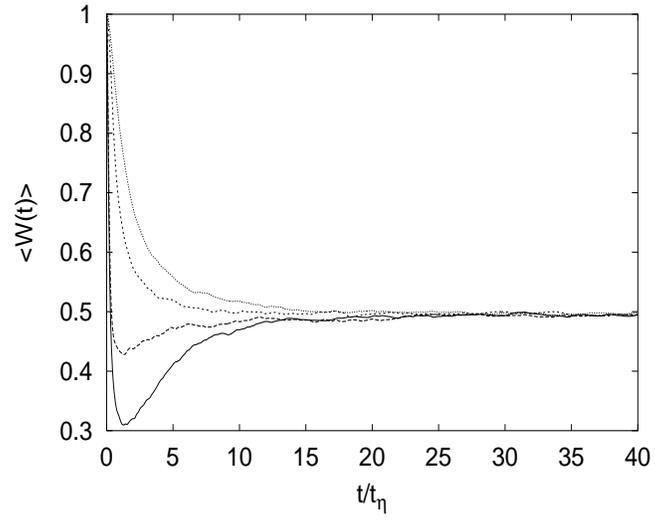}
\end{center}
\caption{Time evolution of $\langle w\rangle$}
\label{fig_2}
\end{figure}
\begin{figure}
\begin{center}
\includegraphics[height=7cm,width=9cm]{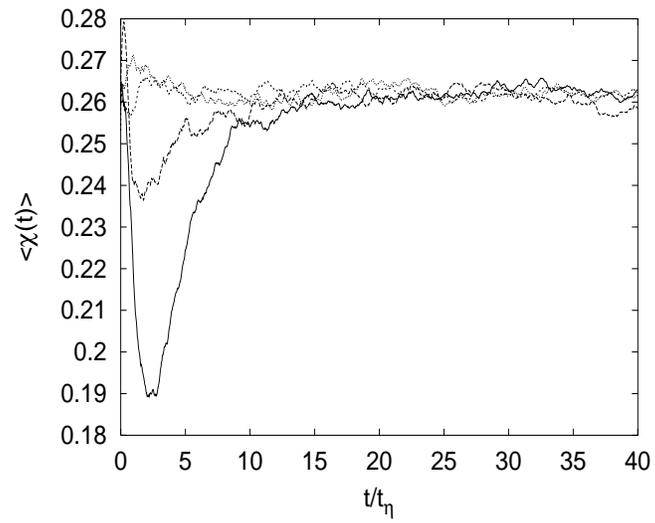}
\end{center}
\caption{Time evolution of $\langle \chi\rangle$}
\label{fig_3}
\end{figure}

The behaviors as functions of time of the mean values of $\langle R
\rangle$ (see Fig. \ref{fig_1}), $\langle w \rangle$ (see
Fig. \ref{fig_2}) and $\langle \chi \rangle$ (see
Fig. \ref{fig_3}), as well as the evolution of the PDFs of
$R$, $w$ and $\chi$ (see Fig. 4) computed from KS are very
close to the ones observed experimentally (see Figs. 3-6 of
\cite{patrizia}). Contrary to the laboratory experiment, where the
spatial confinement of the setup induced a saturation of the growth of
the radius of gyration, $\langle R^2 \rangle$ grows like $t$ at very
long times.  The distribution of sizes, $R$, as well as the
distribution of $w$ and $\chi$ are Gaussian in this regime.  The
relaxation of the peak of probability for $\chi \approx 0$ is faster
in the KS than in the experiment. This small discrepancy is
conceivably due to the large scale limitations of the flow.

At intermediate time scales, when $\langle R^2 \rangle^{1/2}$ is in
the inertial range, the mean values and the distribution of $w$
are very close to the ones observed in \cite{patrizia}.

These results demonstrate that KS reproduces very well the main
properties of the evolution of three Lagrangian particles. A similar
conclusion was reached in \cite{NicolVas2002} by comparing laboratory
and KS results in two dimensions, and in \cite{MkVas99} by comparing
DNS and KS results in $3$-dimensions. KS is thus a potentially very
useful tool both for fundamental studies, and for dispersion studies
in a more applied context.

\section{Predictions of KS in the large Reynolds number
limit}\label{large_Reynolds_limit}

The results of the previous subsection demonstrate that the KS model
reproduces quite satisfactorily the laboratory results concerning the
evolution of three particles in a turbulent $2$-dimensional flow. We
now investigate the large Reynolds number limit with the help of the
KS. This is achieved by increasing the ratio of the largest to the
smallest scale, $L/\eta$ irrespective of initial conditions.

A previous study \cite{NicolVas2002} has shown that the separation
between two particles in the inertial regime shows sizable differences
compared to the famous Richardson's scaling, according to which the
separation grows according to $\langle R^2 \rangle \propto \epsilon
t^3$ irrespective of initial conditions. Specifically a strong
dependence of two-particle dispersion statistics on  the initial
separation the particles was found \cite{NicolVas2002}.  In the case
of three particles studied here, it is of obvious  interest to
investigate the statistics of the radius of gyration of the set of
particles.  Also, the existence of a non trivial, time independent
distribution of shapes,  predicted by \cite{Pumir2000} when the
separation between particles is in the inertial range of scales,
remains to be tested. We address these questions in turn, with the
help of KS.

We stress that there is no guarantee that the evolution of $3$
particles is correctly predicted by the KS with $p=5/3$ when the ratio
$L/\eta$ becomes large. In spite of this uncertainty, the numerical
results presented here are intrinsically interesting since the KS flow
shares with real turbulence a number of important properties.

The evolution of three particles has been followed numerically, using
several KS flows, corresponding to different Reynolds numbers, or
equivalently, to different values of $L/\eta$  ($L/\eta =1691$,
$3381$, $16909$). Our runs, with the values of the parameters
characterizing the flows, are listed in Table 1.

The results of these simulations (Fig. 5, corresponding to $L/\eta=1691$ and
Fig. 6, corresponding to $L/\eta=16909$)  show some resemblance with the
small $L/\eta$ case, see section \ref{exp_ks_results}). As the
Reynolds number is increased, the dependence of the variables
describing geometry, $\langle I_{2}\rangle$, $\langle W\rangle $, and
$\langle\chi\rangle$ becomes weaker, both as a function of the initial
size of the triangle and as a function of time.  The values of these
variables is always  significantly different from the Gaussian values.

The PDF of $R$, $w$ and $\chi$ at the value $L/\eta = 1691$ also
show trends which are similar to the ones observed  at smaller
values of $L/\eta$ (Fig. 7).  We have not followed
particles long enough to see the Gaussian distribution of shapes
at very large values of $L/\eta$; the study has been restricted to
the non trivial inertial range. In this range of scales, the
variation of the PDF as a function of time is considerably weaker
than observed in section \ref{exp_ks_results}, and in this sense,
the results suggest that one may be getting close to the self
similar shape distribution predicted in \cite{Pumir2000}.  A
visible deformation of the PDF of $R$ can be observed as a
function of time, reflecting the fact that at the later times,
large excursions in the radius of gyration are getting close to
the value of the largest available scale, $L$.

Although still finite, the value of $L/\eta$ for run $13$ is
significantly larger than it is in any engineering,  industrial or
laboratory flow. Even so, our numerical results show
persistent differences with the picture of a simple truly self
similar regime.

In all the reported DNS, in $2$ \cite{celani2000} and in $3$ dimensions 
\cite{yeung94}, a dependence of
the behavior of $\langle R(t) \rangle$ on the initial separation
$r_{0}$ has been reported. We observe a similar behavior in our KS
calculation. Fig. 8 shows that the $t^{3}$ Richardson
regime is never really reached. Instead, a continuous dependence of
the variation of $\langle R^2 \rangle/t^3$ on the initial
separation $r_0$ is observed. This effect is clearly seen, even at our
largest Reynolds number. This behavior suggests that the set of three
particles always remembers its initial condition, which represents a
departure with respect to the Richardson prediction. Also, at a fixed
value of the  initial size of the triangle, a non trivial power seems
to emerge as the Reynolds number increases. A similar behavior was
observed for the separation between two particles \cite{NicolVas2002}.

To investigate further this effect, we consider the PDFs of
the radius of gyration, $R$. Fig. 9 shows a
superposition of the PDFs of $R/\langle R \rangle$ for three
different initial values of $r_0/\eta = 0.1,\ 1.0\ 5.0 $, and at
different values of time. The distribution of the large values of
$R/\langle R \rangle$ are expected to be independent of time and
Fig. 9 does not disprove this expectation. However,
the distribution at small values of $R/\langle R \rangle$ seems to
vary throughout the entire evolution.

To compare different values of $r_0$, we have plotted the PDFs of
$R/\langle R \rangle$ corresponding to two different values of
$r_0$, but with similar values of $\langle R \rangle$ (Fig. 10, 11). 
For large enough times the PDF of
$R/\langle R\rangle$ seem to collapse at large values of
$R/\langle R \rangle$ within statistical errors. However,
serious deviations are observed at small values of  $R/\langle R
\rangle$. In particular the peaks of the PDF at very small
values of $R/\langle R \rangle$ are much sharper at smaller
values than at larger values of $r_0$.

These results suggest that the distributions of $R$ are influenced
throughout the entire evolution by the initial value $r_0$, insofar as
a significant number of triplets do not really separate, and remain at
a value $R \sim r_0$. In the case of two particles this behavior has
indeed been observed by Jullien et al. \cite{jullien99} in the
laboratory for low Reynolds number flows, by Fung et al. \cite{Vas98} and
Nicolleau et al. \cite{NicolVas2002} in high Reynolds number KS
simulations. In this way, the evolution depends in an essential way on
the value of $r_0$.

The evolution of other geometric quantities, such as $\langle w
\rangle$, $\langle \chi \rangle$ reflects to some extent  the behavior
described above (see Fig. 5, 6). Indeed, the lack of exact self
similarity observed in the evolution of the radius of gyration, $R$,
shows that the prediction of a truly time independent shape
distribution is at best valid at Reynolds numbers impossible to
attain. Although this prediction might constitute a good first order
approximation, which becomes better as the Reynolds number increases,
Fig. 6 shows that the mean  values of $w$, $I_{2}$ and $\chi$ {\it do }
vary with time, even at $L/\eta = 16909$.  In addition, a systematic
variation with $r_0$ is seen in Fig. 5, 6.  The distributions are
observed to remain non Gaussian as long as $\langle R \rangle$ remains
in the inertial range, and do correspond to a higher probability of
observing elongated objects, as anticipated in the stochastic model
proposed in \cite{pumir,Pumir2000}. In the light of the KS results, the
stochastic model correctly predicts the main qualitative feature (the
increased probability of elongated objects), but it doesn't
incorporate the lack of self-similarity of multi-particle diffusion
observed in KS. As we discuss in the next paragraph, memory effects
relating to this lack of self-similarity are observed in laboratory
experiments \cite{jullien99,patrizia}. They are also observed in KS
because, unlike stochastic models, KS incorporates the persistence of
flow structures. Whether this lack of self-similarity persists in the
laboratory and in nature at extremely high Reynolds numbers remains an
open question.

The results obtained so far suggest that the lack of self-similarity
in the evolution of the radius of gyration, $R$, is due to the fact
that particles stay together, at a distance of order $r_0$ with a high
probability.  Even if the radius of gyration grows to a value $R\gg
r_0$, it was observed in \cite{patrizia} that two particles of the
triplet can remain close to each other, with a large probability. We
interpret the lack of a stationary distribution of shapes to reflect a
similar cause : when one particle of the triplet separates from the
two other ones, which remain at a mutual distance $\sim r_0$, a very
elongated shape is created.  The relaxation of the distribution of
shapes towards a stationary distribution will depend on how the two
particles that are close together eventually  separate. Our
observations suggest that, the smaller $r_0$, the longer particles will
stay together, hence, the longer it will take the transient to relax.
This lack of self-similarity is absent in stochastic models, such as
the one proposed by \cite{pumir}, and is consistent with the view that
coherent streamline structures are persistent enough to cause a
dependence of turbulent diffusion on initial conditions.

\section{Effects of persistence of the flow structure}\label{ks_new_shape}

The KS model allows us to modify some of the characteristics of the
advecting flow, both spatially and temporally. This is achieved
by modifying the parameters $\lambda $ and $p$. The purpose of this section
is to investigate the effect of changing the spatio-temporal structure
of the flow, and in this way, to gain insight into the
mechanisms involved in multi-particle dispersion.

We firstly change the temporal structure of the flow by varying the
persistence parameter $\lambda$ (see Fig. 12 ). This controls
how fast the streamlines of the flow are jittered in comparison to the
relevant eddy turn over time at the corresponding scale. This
jittering makes the particles in the flow to be rapidly swept from one
streamline to the other. Since in KS there is no interaction among
modes of the velocity field, this jittering mimics the sweeping
effects that are present in a real flow field.
The minima of $\langle I_{2}(t)\rangle$ and  $\langle w(t)\rangle$
increase with increasing unsteadiness parameter $\lambda$ within the
inertial range of time scales which means triangles are less elongated
for larger values of $\lambda$ (Fig.12). This happens because the
paths of neighboring particles decorrelate faster for larger values of
$\lambda$ and triangles quickly forget their memory of the initial
state. Hence increased values of $\lambda$ should cause the triangle
shape parameters to relax faster to their corresponding Gaussian
values.

Secondly we change the spatial structure of the flow field by changing
the energy spectrum i.e. changing the exponent $p$ in $E(k)\sim k^{-p}$
(see Fig. 13 ). This has the effect of changing the density
of straining regions in the flow field \cite{Vas98,davila02} thereby
modifying the separation mechanism of particle pairs and clusters. The
minimum of $\langle I_{2}(t)\rangle$ and  $\langle w(t)\rangle$
decreases with increasing $p$ ($E(k)\sim k^{-p}$) within the inertial
range of time and scales. This means that the clusters are more/less
elongated during the inertial range of times when $p$ is made
larger/smaller. An explanation of this effect can be given in terms of
randomness: as $p$ increases there is less energy in the smaller
scales of the turbulence which may mean less randomness leading to
clusters remaining more elongated during the inertial range of times.
In the context of KS as $p$ increases there is indeed less energy in
the small scales leading to smaller unsteadiness frequency
$\omega_{n}\sim k_{n}^{(3-p)/2}$ and therefore less randomness and
more elongated clusters. However, a more searching explanation should
invoke the spatio-temporal flow mechanism causing cluster
elongation. One such mechanism already proposed in the literature
\cite{Vas98,FHMPerk92} is based on persistent effects of straining
regions. The spatial density of straining regions decreases as $p$ is
made larger \cite{Vas98,davila02}.

However, the simple calculation, presented in the Appendix, indicates
that the average straining rate per straining region increases faster
than the number of such regions decreases when $p$ is made
larger. Provided that the effect of increased strain rate per strain
region overwhelms that of the decreased number of such regions, then
the same conclusion is reached: clusters should remain more/less
elongated during the inertial range of times when $p$ is made
larger/smaller.


\section{Conclusions}\label{ks_shape_con}

We have investigated the Lagrangian shape dynamics and the
corresponding statistics of multiple particles namely of three
particles advected by a two dimensional turbulent flow. We have used
kinematic simulation (KS) to generate a turbulent velocity field and
follow numerically sets of three particles in this flow field. The
results of the simulation have been compared with the results of a two
dimensional experiment \cite{patrizia}. We have identified a mechanism
for the shape evolution of three particles depending on the underlying
flow structure and the effect of persistence of these structures on
the statistics of these shapes.

Two regimes with well-characterized distributions have been identified
in our simulation with KS. These regimes have also been observed in
the two dimensional experiments of Castiglione
et al. \cite{patrizia}. Two different regimes can be identified,
according to the fact that the mean separation is large compared to
the largest scale (diffusion regime) or in the inertial regime.
The diffusion regime is characterized by a Gaussian shape distribution.

In the inertial regime the scale $\eta^{2}\ll \langle R^{2}\rangle \ll
L^{2}$, we have observed what might be described as a trend, at best,
towards a Richardson's law $\langle R(t)^{2}\rangle\propto t^{3}$. But
the appearance of this regime was found to be dependent on the initial
scale or size $r_{0}$ of the triangles. This has also been observed in
the experiment \cite{patrizia}.

The temporal evolution of $\langle I_{2}(t)\rangle$ , $\langle
w(t)\rangle$ and $\langle\chi(t)\rangle$ match with the experimental
results of \cite{patrizia}. KS predicts the correct temporal evolution and the
distribution functions of the above quantities. Monte Carlo simulation
can do as well except for the PDF of $\chi$ \cite{patrizia}.

It is found that the clusters are more/less elongated during the
inertial range of times when $\lambda$ is made smaller/larger (Fig. 12).
The reason for this result must be that the
persistence of the straining action of the flow is diminished when the
flow is made more unsteady by increasing $\lambda$.

It is also found that the clusters are more/less elongated during the
inertial range of times when $p$ is made larger/smaller. An
explanation of this effect can be given in terms of randomness: as $p$
increases there is less energy in the smaller scales of the turbulence
which may mean less randomness leading to clusters remaining more
elongated during the inertial range of times. However, we also discuss
a more searching explanation which invokes the straining mechanisms
causing cluster elongation.

The dependence of the shape of clusters on the initial separation
between marked fluid elements is clearly demonstrated by the PDF of
the radius of gyration of particle clusters with different initial
separations not collapsing within the inertial range (see Figs.10,11).
Our KS numerical experiments indicate that clusters tend to have
memory of the initial state even when the turbulence has an extremely
wide inertial range of more than four decades.

This result is in agreement with the observed dependence \cite{NicolVas2002}
on the initial pair separation of the apparent power law governing the
growth of inter-particle distances. If these effects are transient and
due to a finite range of inertial scales, then our results indicate that they
might only disappear at extremely high Reynolds numbers. However,
the possibility should also be retained that these effects are not
finite range transient effects, and are caused, instead, by the
persistence of spatial flow structures at all scales, assuming this
persistence remains at asymptotically high Reynolds numbers. 

\section*{Acknowledgements}

We acknowledge the European contract HPRN-CT-2002-00300 and ECTMR
network on intermittency in turbulent systems. MAIK acknowledges
support from the Cambridge commonwealth trust (CCT), overseas research
scheme (ORS), DAMTP and Wolfson College, Cambridge. JCV acknowledges
financial support from the Royal Society, UK and the Hong Kong
research grant council (project number HKUST6121/00P). AP acknowledges
support from IDRIS, France.

\section*{Appendix}

Assume we are given the spectrum $E(k)=E_{0}L (kL)^{-p}$ defined in
the range $1/L\leq k\leq 1/\eta$ of an isotropic turbulence. The mean
square straining rate $\langle \left({\partial u}/{\partial
x}\right)^{2}\rangle $ is proportional to $\int_{1/L}^{1/\eta}
k^{2}E(k)\; dk$.
Substituting the form of the spectrum
and integrating we get for $p<3$, 
\begin{equation}
\left\langle \left({\frac{\partial u}{\partial x}}\right)^{2}\right\rangle\sim{\frac{E_{0}}{(3-p)L^{2}}}\left({\frac{L}{\eta}}\right)^{3-p}.
\end{equation}
In KS, the number density of straining stagnation points
decreases with increasing $p$ \cite{Vas98} and \cite{davila02}
calculated the following scaling relation:
\begin{equation}
n_{s}\sim \left({\frac{L}{\eta}}\right)^{D_{s}},
\end{equation}
where $n_{s}$ is the number density of straining stagnation points and
$D_{s}$ is the fractal dimension of the spatial
distribution of these points in the flow. In 2-D KS,
$D_{s}=3-p$ \cite{davila02}. Hence the number of straining stagnation
points per unit area decreases with increasing $p$ in our KS but the
mean square strain rate per straining stagnation point, i.e. \(\langle
({\frac{\partial u}{\partial x}})^{2}\rangle/n_{s}\), scales like
$E_{0}/(3-p)L^{2}$. This implies that, although the number density
decreases with increasing $p$, but the mean strain rate per straining
stagnation point becomes stronger, which is the reason behind the
decrease of the parameters $\langle I_{2}\rangle$ and $\langle
W\rangle$ with increasing $p$.


\begin{figure}
\includegraphics[height=4cm,width=4.5cm]{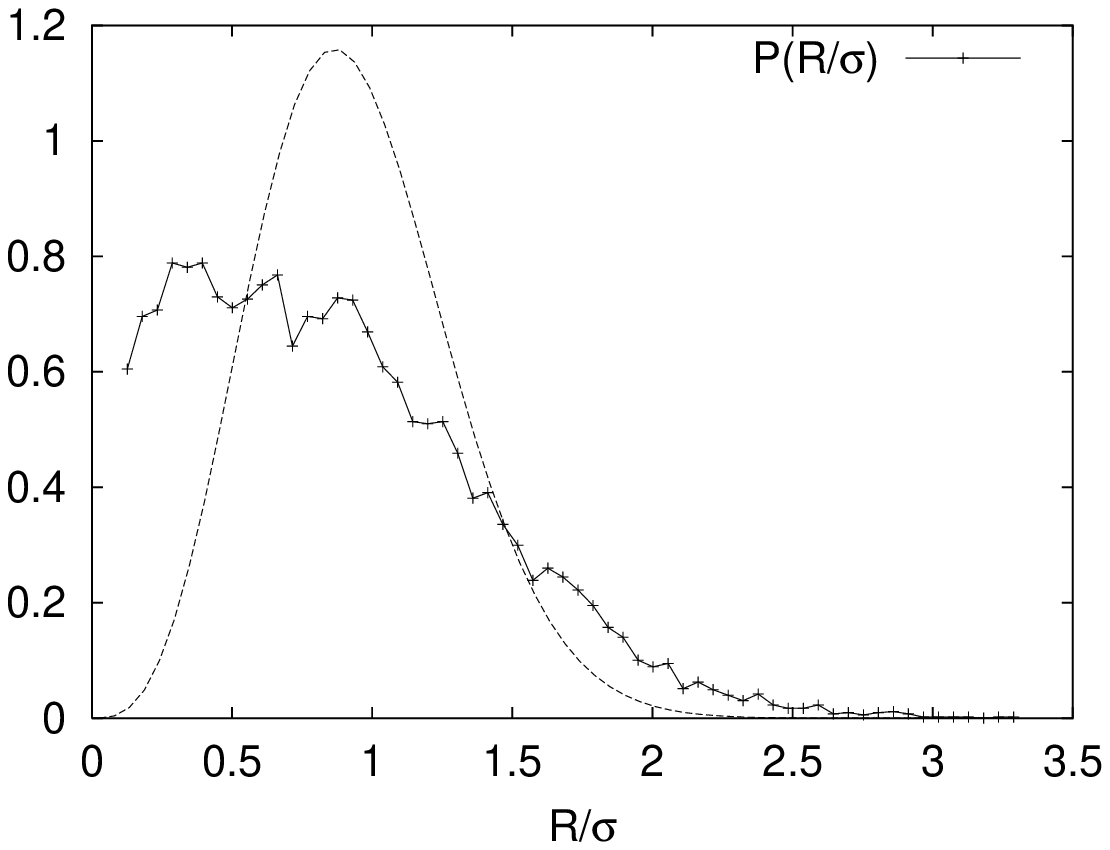}
\includegraphics[height=4cm,width=4.5cm]{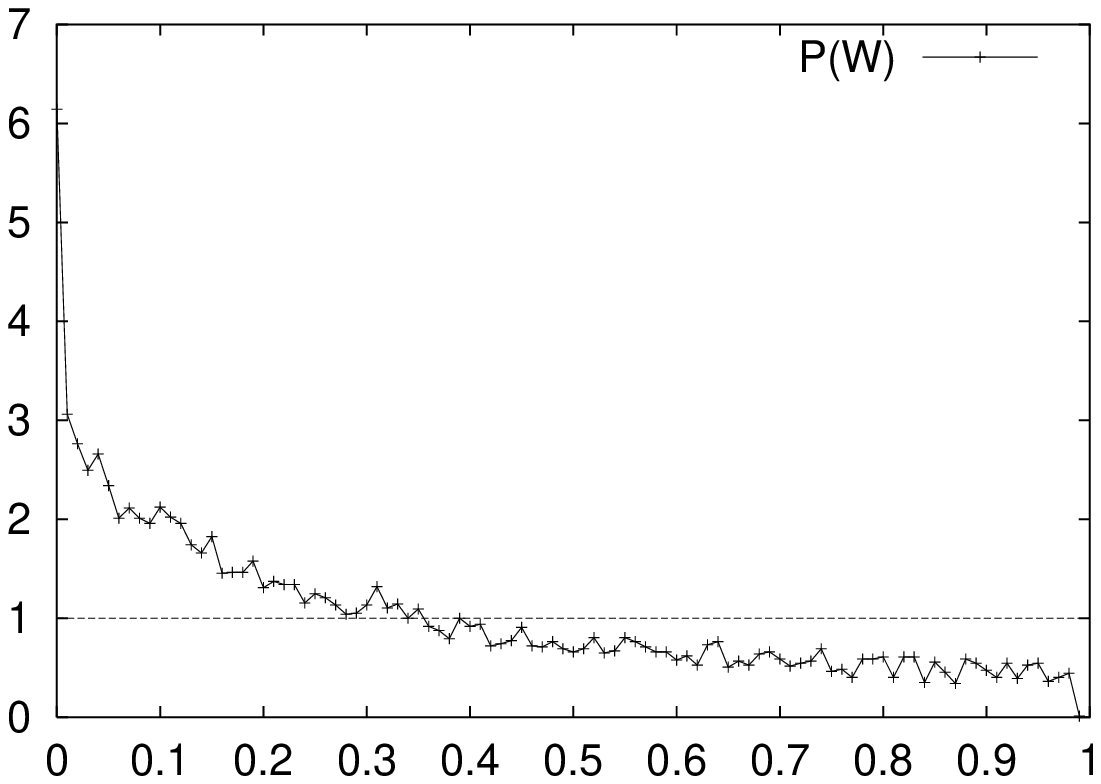}
\includegraphics[height=4cm,width=4.5cm]{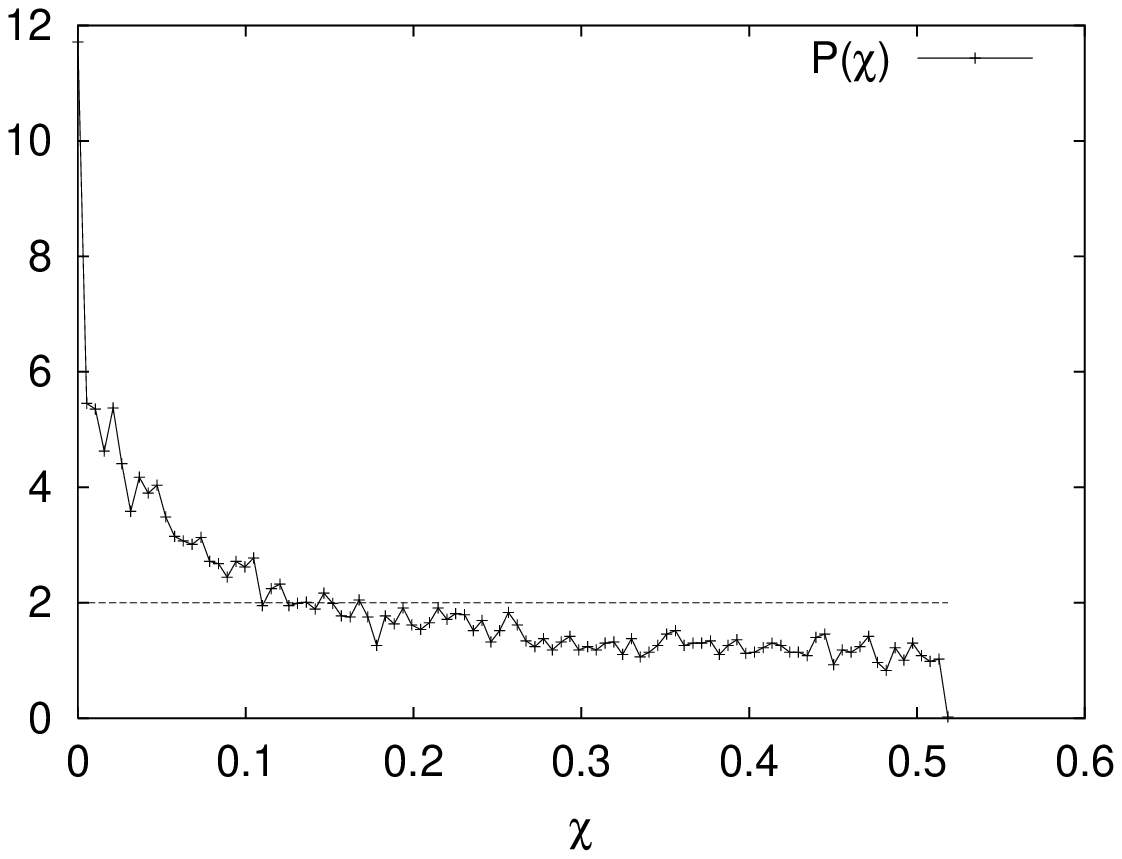}\\
\includegraphics[height=4cm,width=4.5cm]{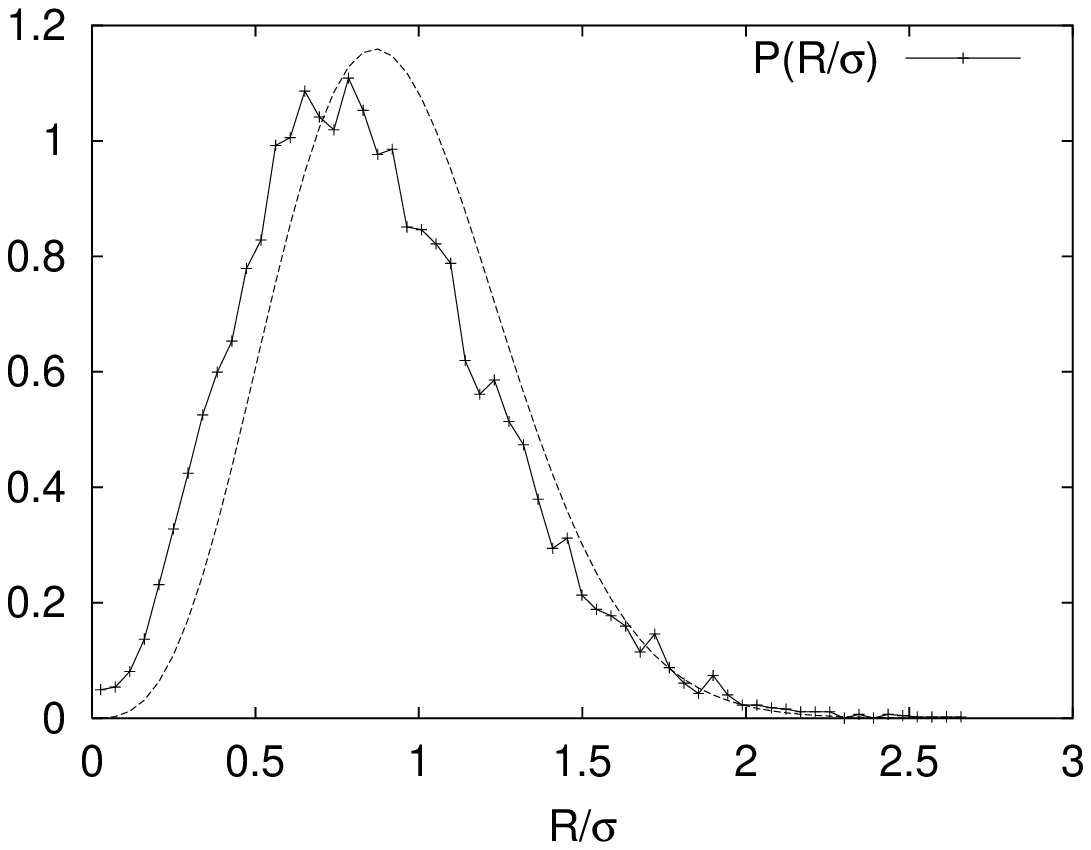}
\includegraphics[height=4cm,width=4.5cm]{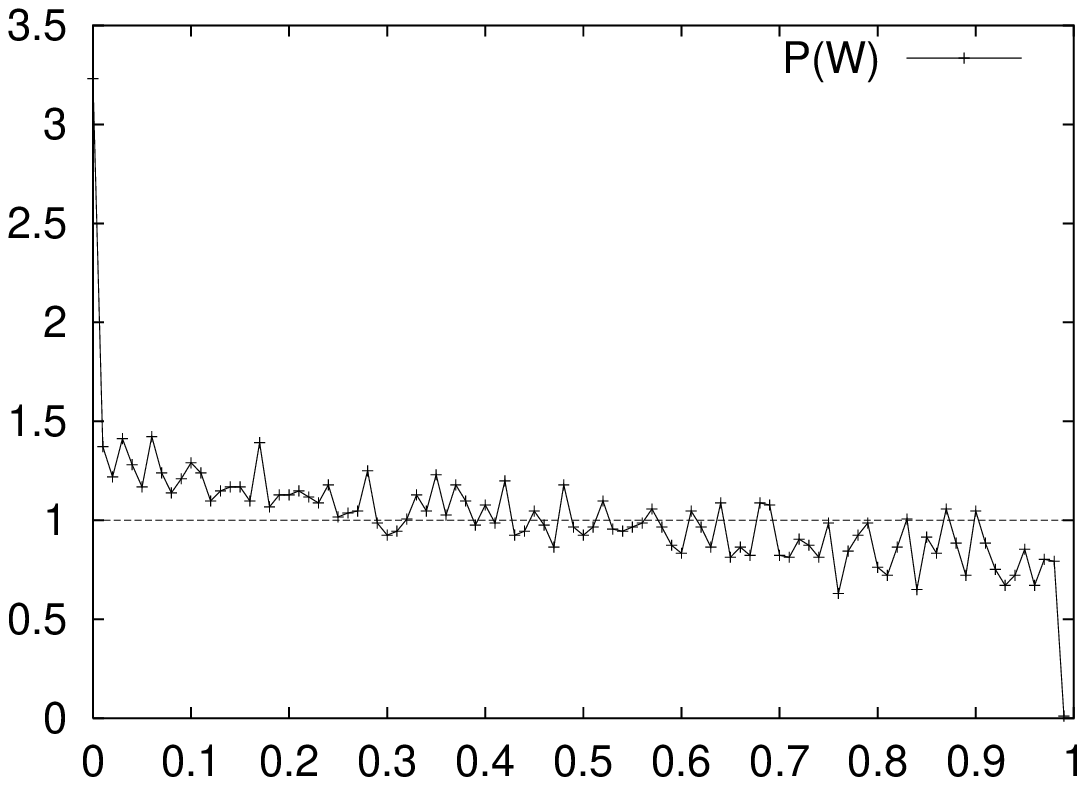}
\includegraphics[height=4cm,width=4.5cm]{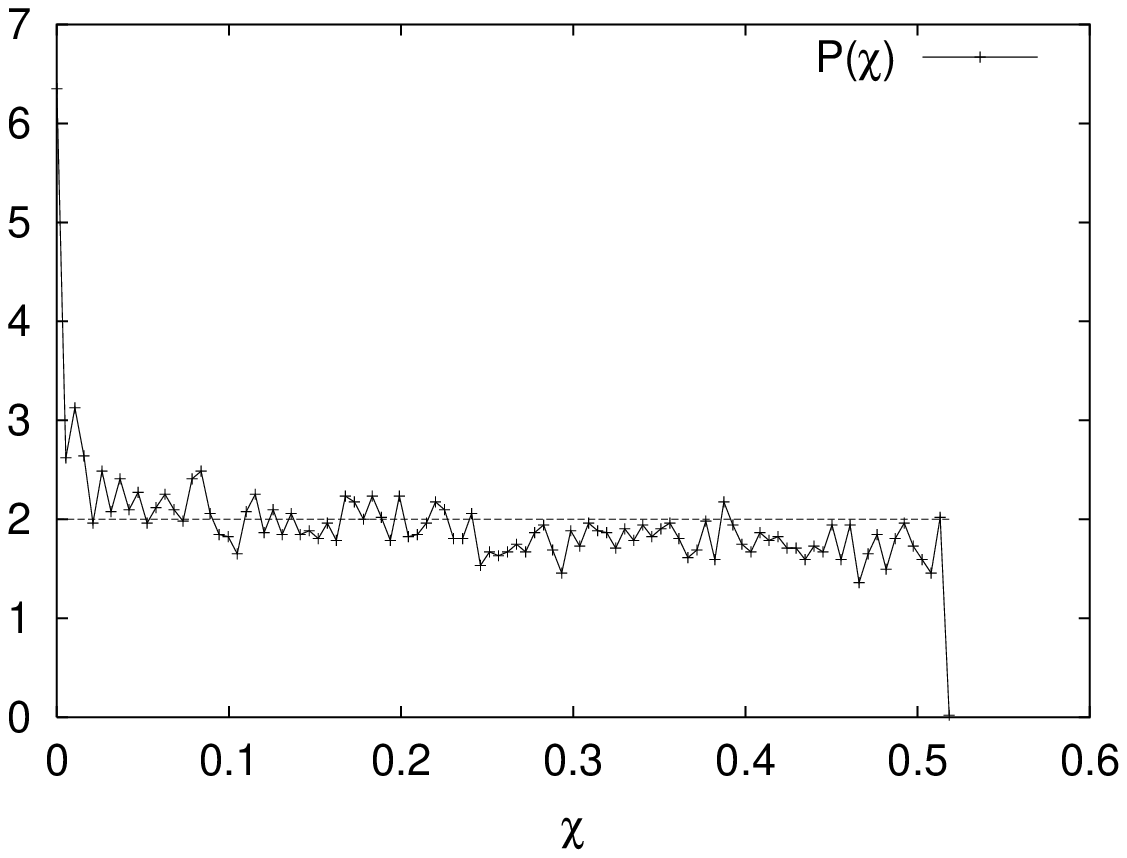}\\
\includegraphics[height=4cm,width=4.5cm]{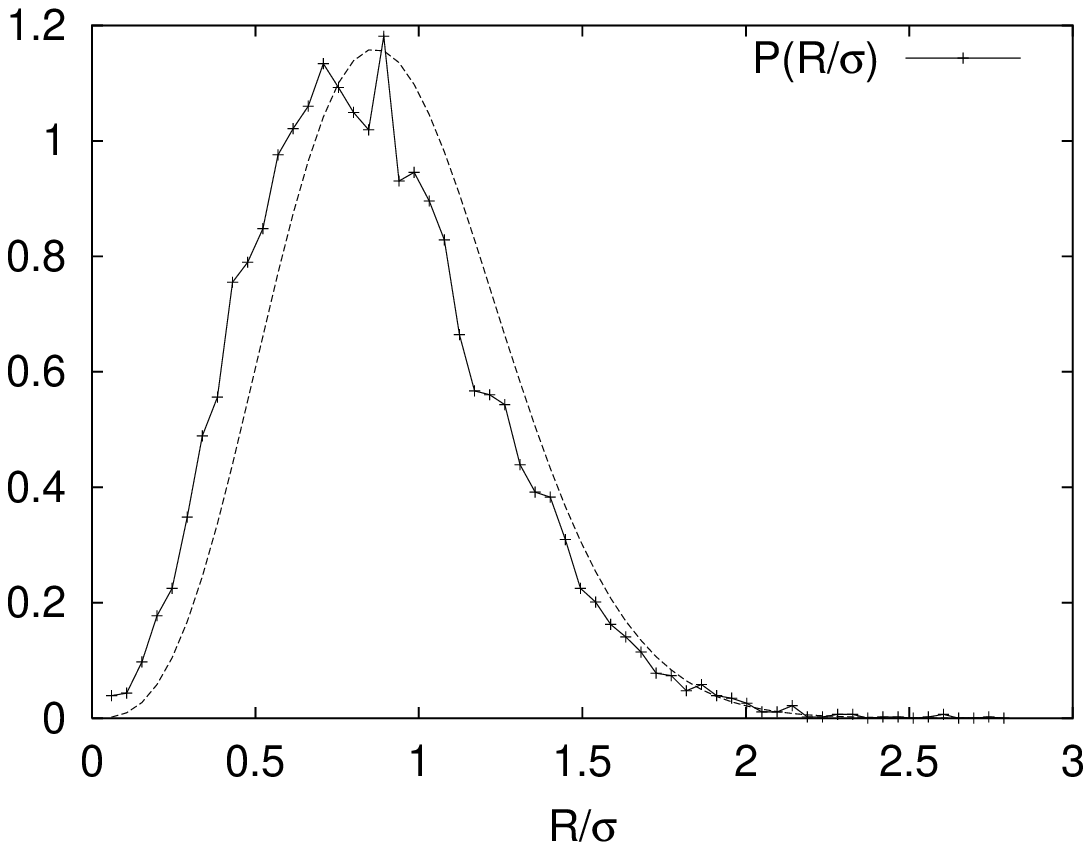}
\includegraphics[height=4cm,width=4.5cm]{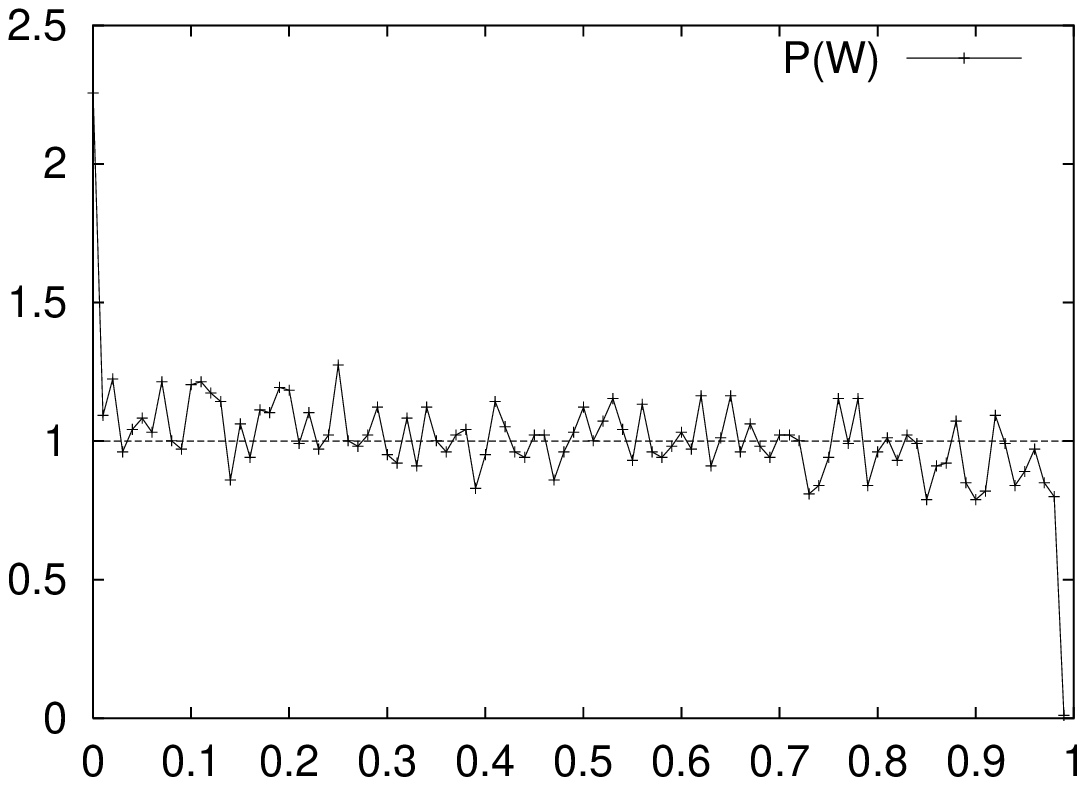}
\includegraphics[height=4cm,width=4.5cm]{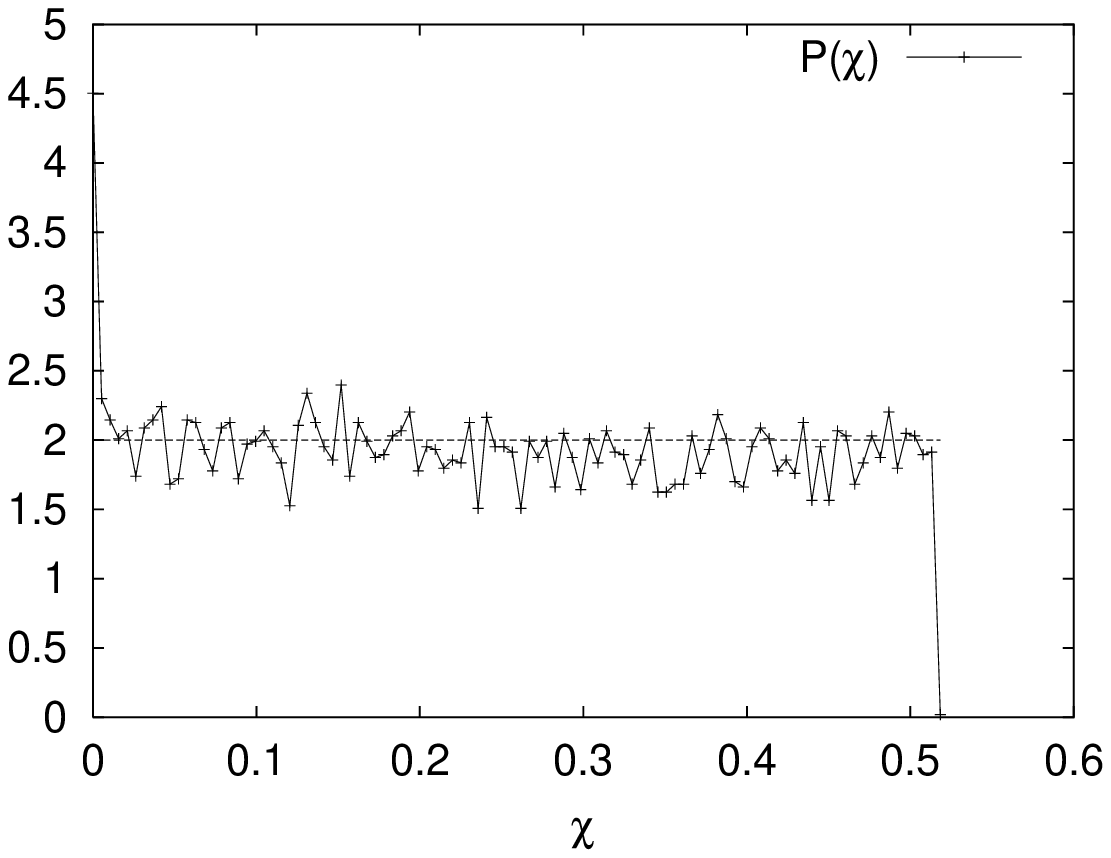}\\
\includegraphics[height=4cm,width=4.5cm]{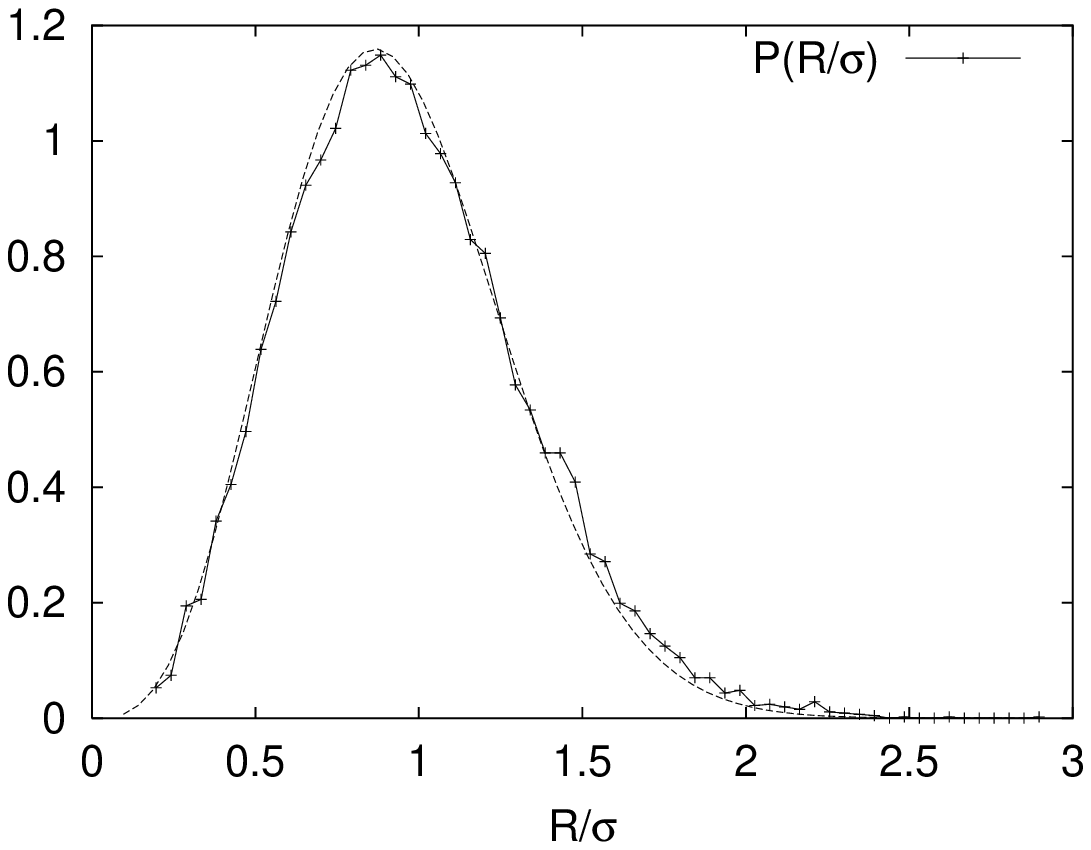}
\includegraphics[height=4cm,width=4.5cm]{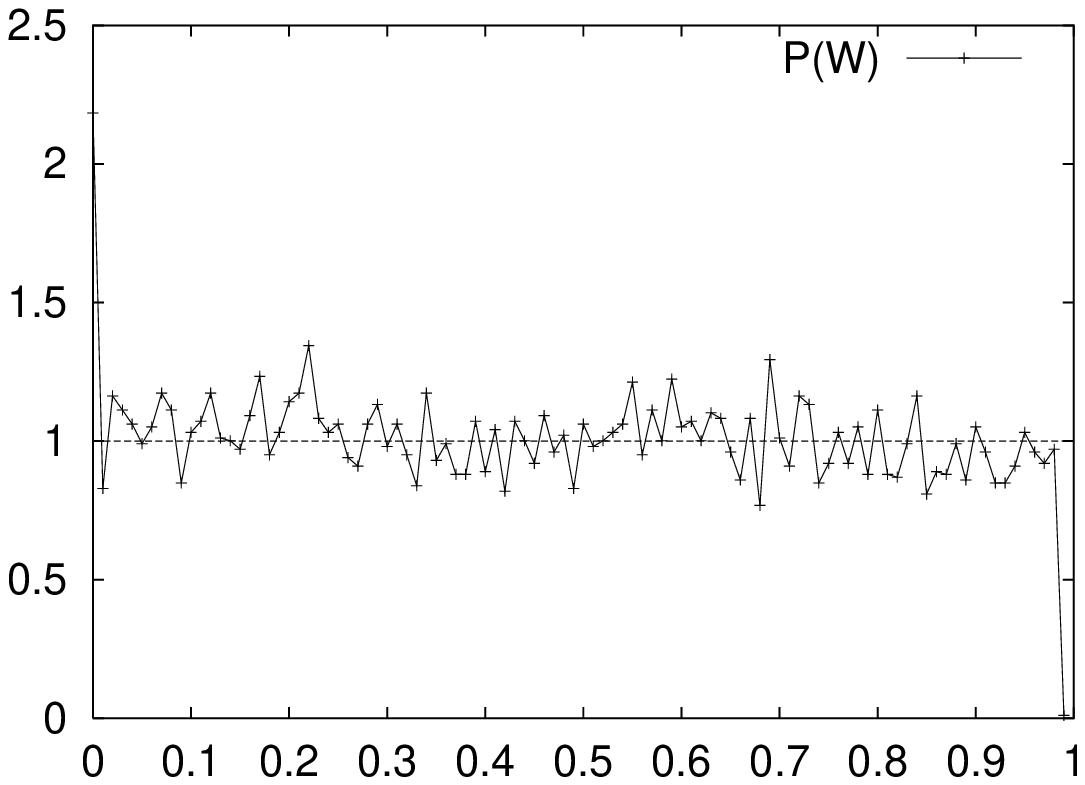}%
\includegraphics[height=4cm,width=4.5cm]{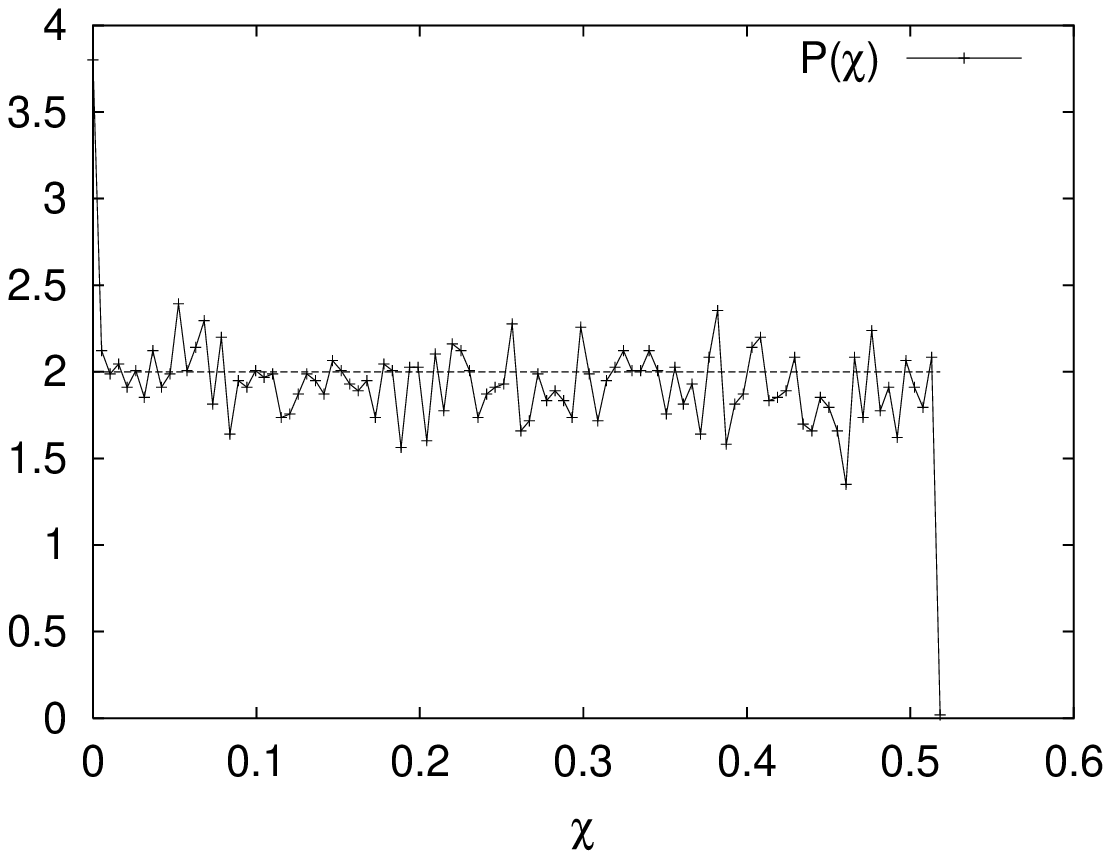}
\caption{\small{Time evolution of the PDF of $R/\sigma$ ($\sigma=\sqrt{\langle
R^{2}\rangle}$), $w$ and $\chi (rad)$ for $r_{0}/\eta=1/6$ in 2D KS, with
$L/\eta=3.67$, $t_{E}/t_{\eta}=2.3$ and $\lambda=0.5$. From top to
bottom the figures are shown at times $t=2 \times t_{\eta}$, $t=6
\times t_{\eta}$, $t=10 \times t_{\eta}$ and  $t=14 \times t_{\eta}$
respectively. The light lines correspond to the Gaussian predictions
$P(R)=8(R/\sigma)^{3}\exp[-2(R/\sigma)^{2}]$, $P(w)=1$ and
$P(\chi)=6/\pi $ [21].}}
\label{fig_4}
\end{figure}
\begin{figure}
\includegraphics[height=5cm,width=7cm]{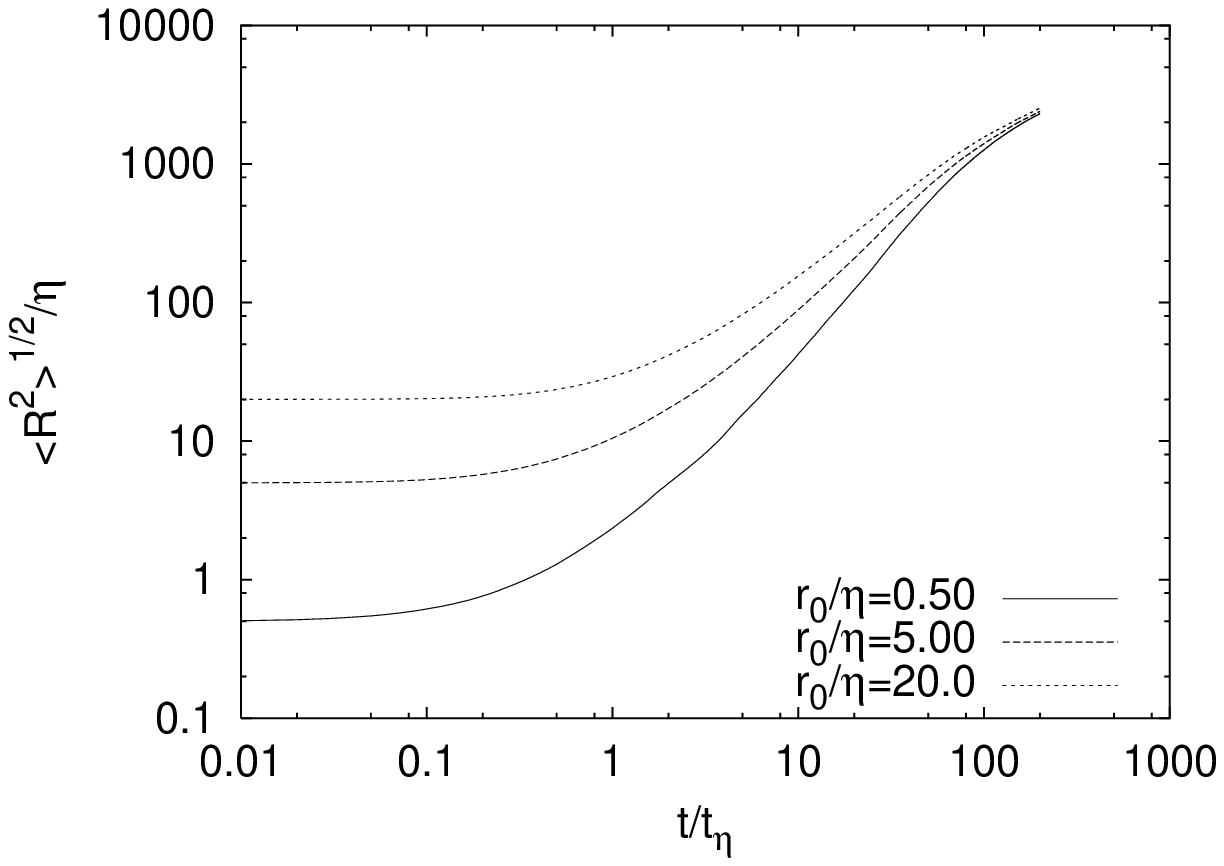}
\includegraphics[height=5cm,width=7cm]{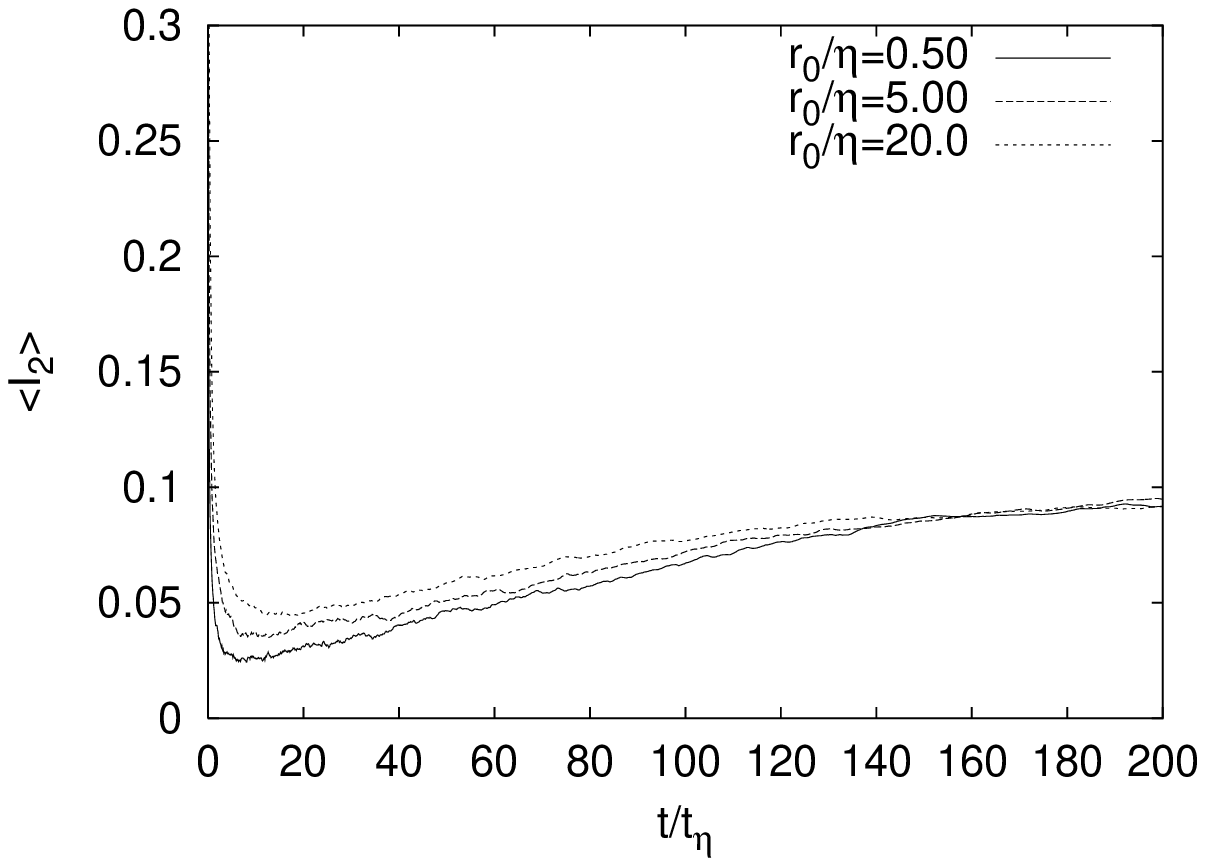}

\includegraphics[height=5cm,width=7cm]{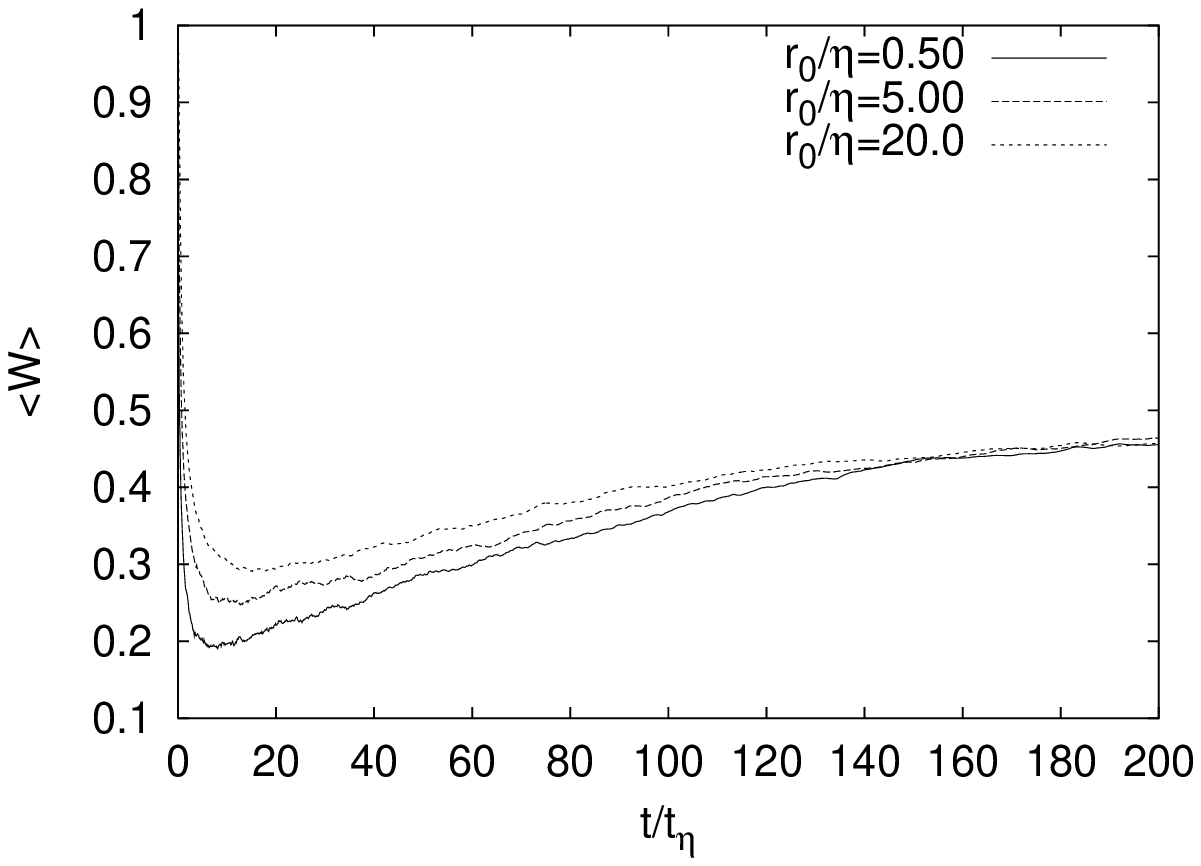}
\includegraphics[height=5cm,width=7cm]{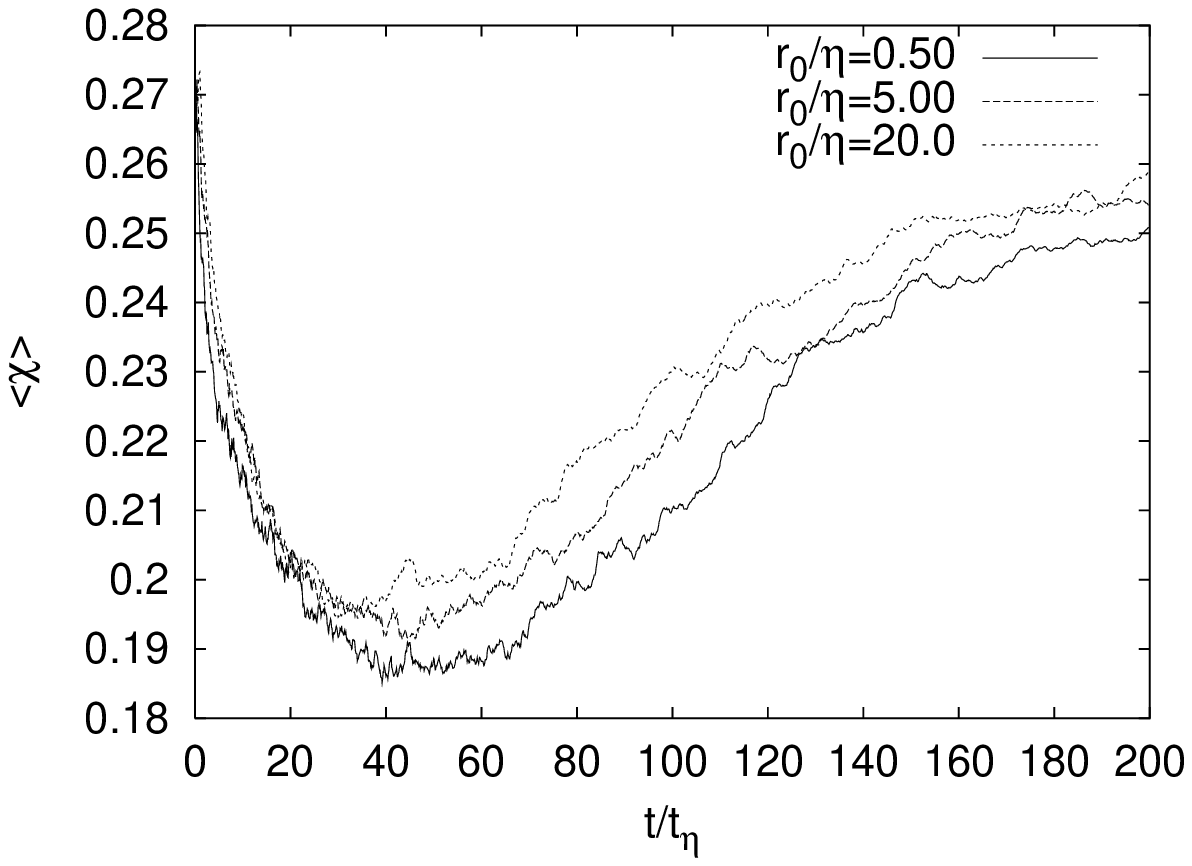}
\caption{\small{Time evolution of $\langle R(t)\rangle$, $\langle
 I_{2}(t)\rangle$, $\langle w(t)\rangle$ and $\langle
\chi(t)\rangle$ produced by Kinematic simulation (KS) in 2D for
$L/\eta=1691 $ and $\lambda=0.5$ with $t_{E}/t_{\eta}=82.1$.}}
\label{KS_WChi_t}
\end{figure}
\begin{figure}
\includegraphics[height=5cm,width=7cm]{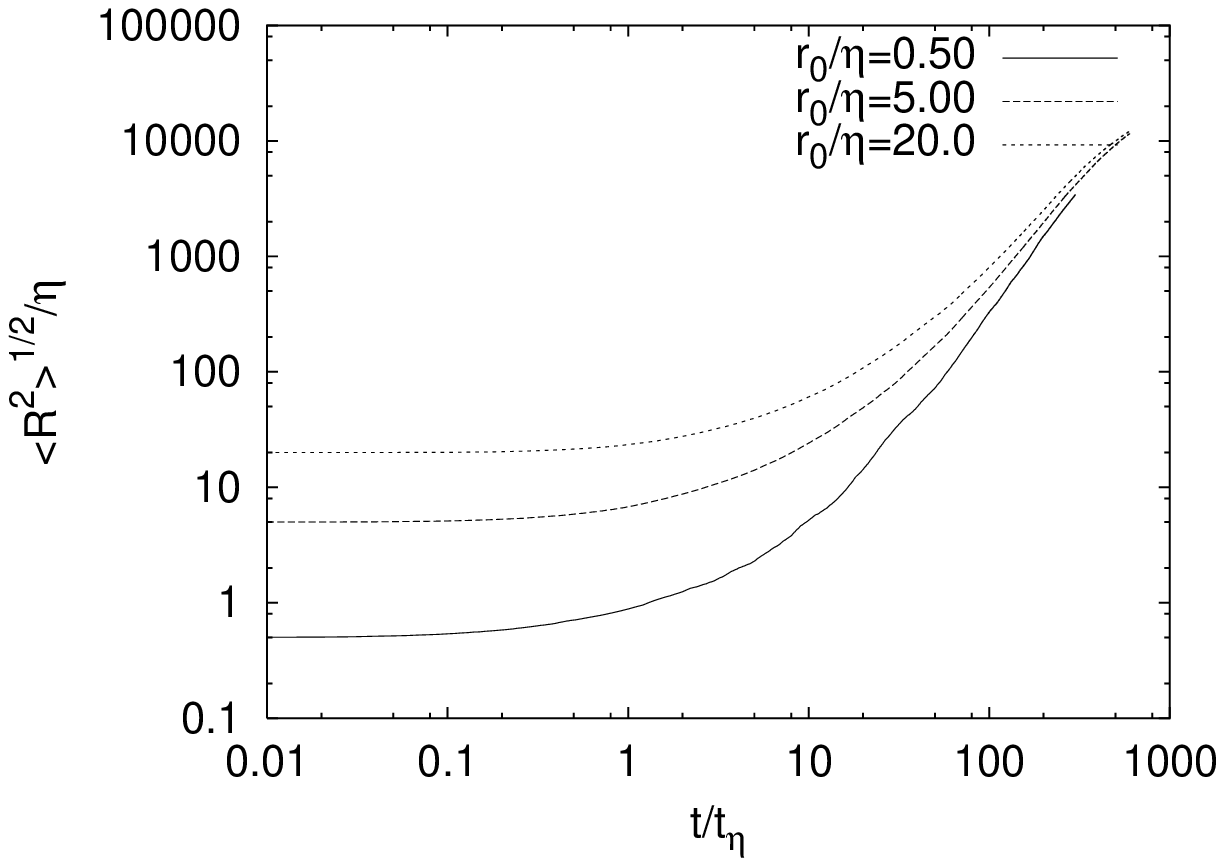}
\includegraphics[height=5cm,width=7cm]{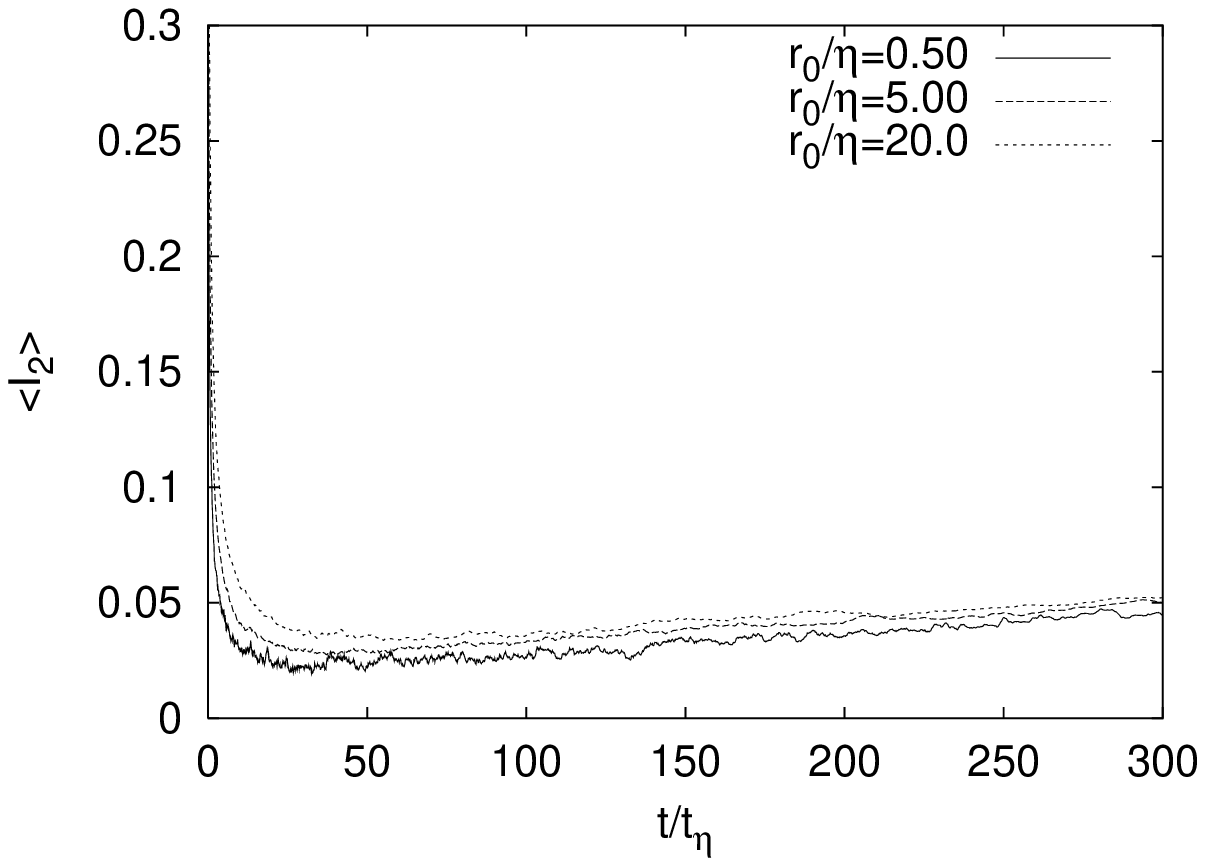}

\includegraphics[height=5cm,width=7cm]{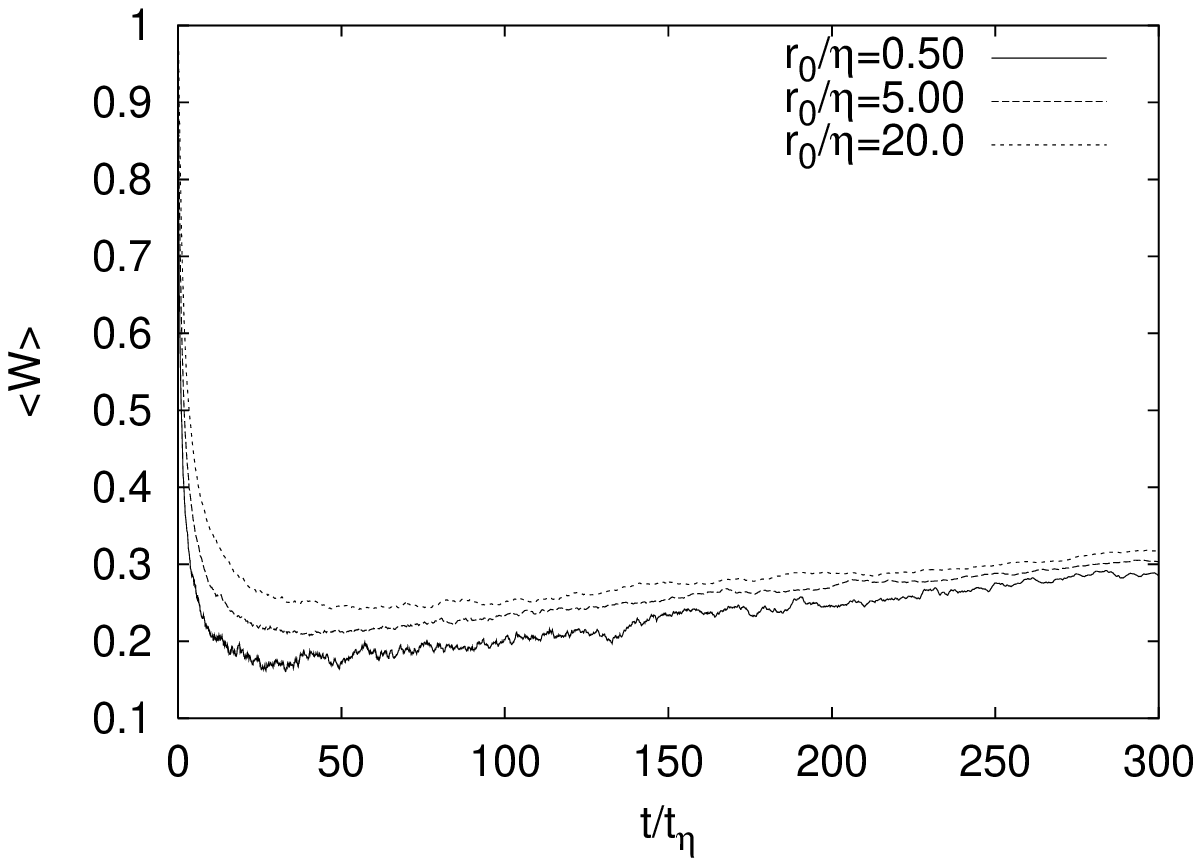}
\includegraphics[height=5cm,width=7cm]{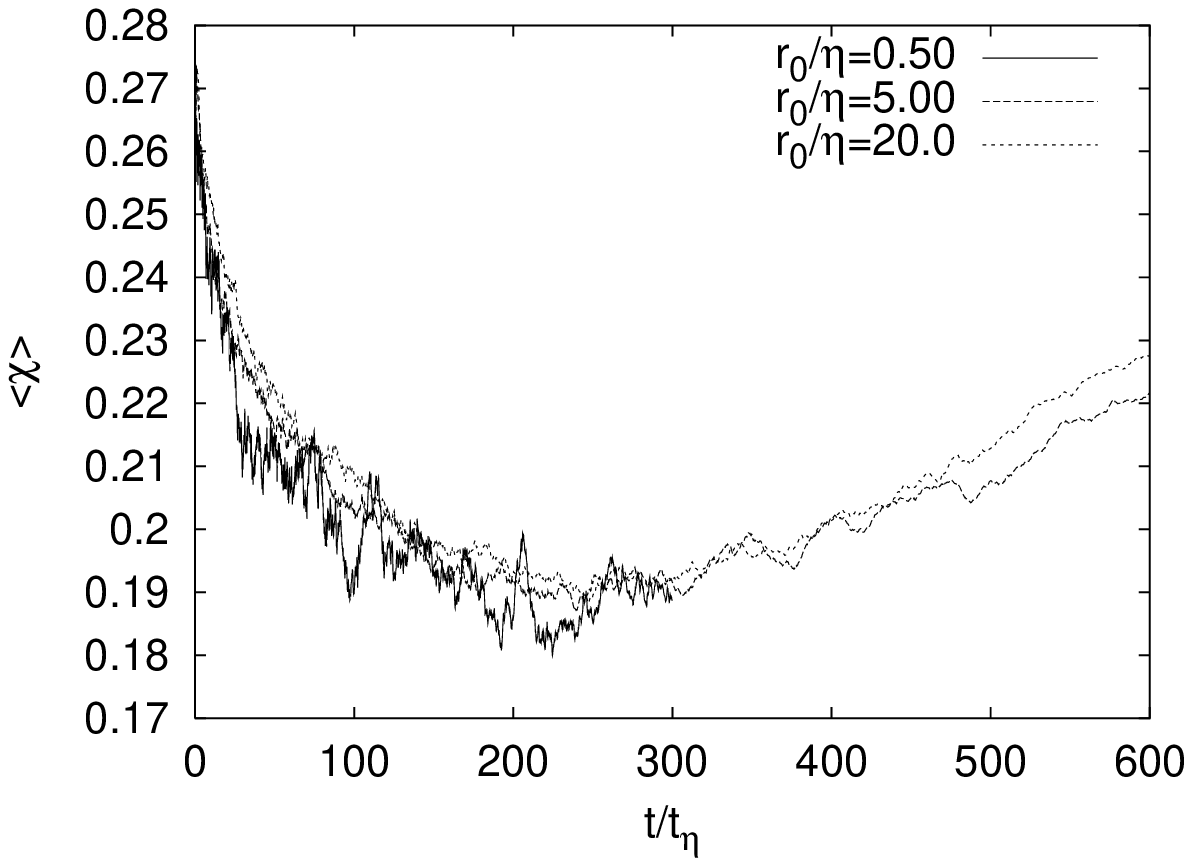}
\caption{\small{Time evolution of $\langle R(t)\rangle$,
$\langle I_{2}(t)\rangle$, $\langle w(t)\rangle$ and
$\langle\chi(t)\rangle$ produced by Kinematic simulation (KS) in 2D
for $L/\eta=16909$ and $\lambda=0.5$ with $t_{E}/t_{\eta}=538.1$.}}
\label{KS_WChi_t_2}
\end{figure}
\begin{figure}
\includegraphics[height=4cm,width=4.5cm]{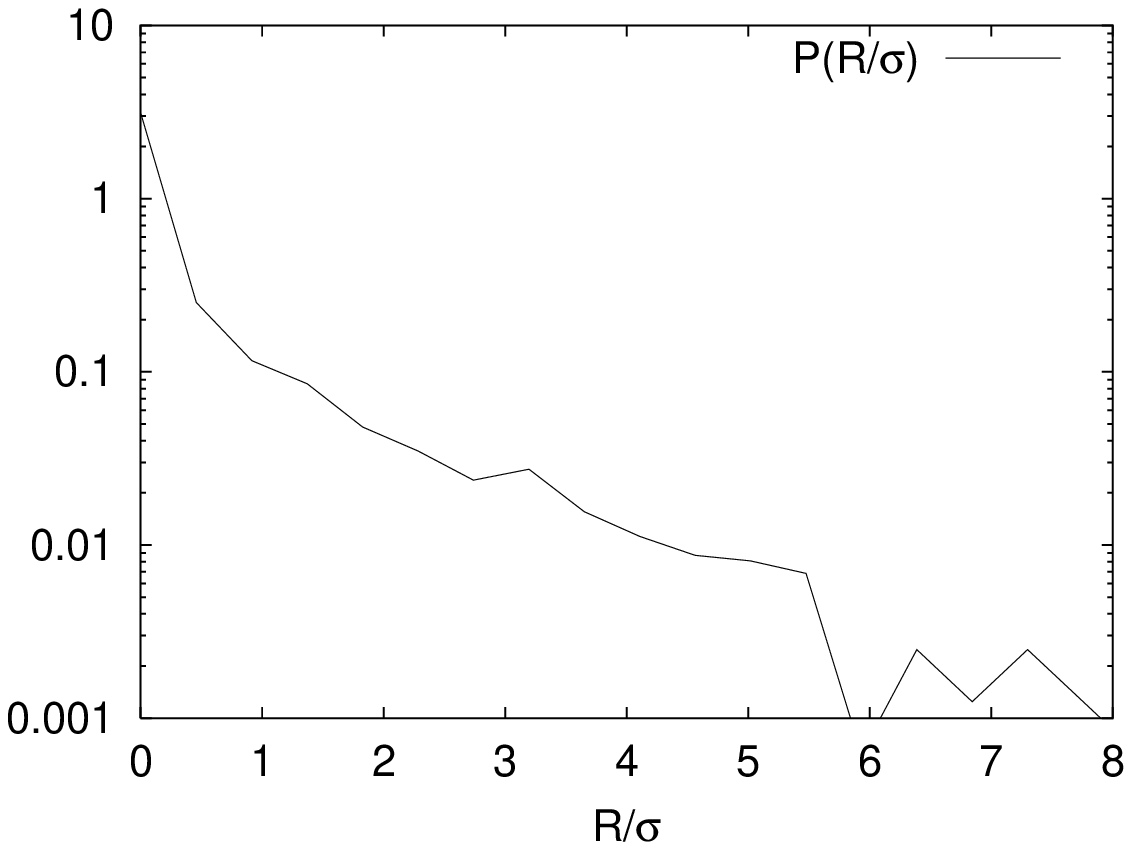}
\includegraphics[height=4cm,width=4.5cm]{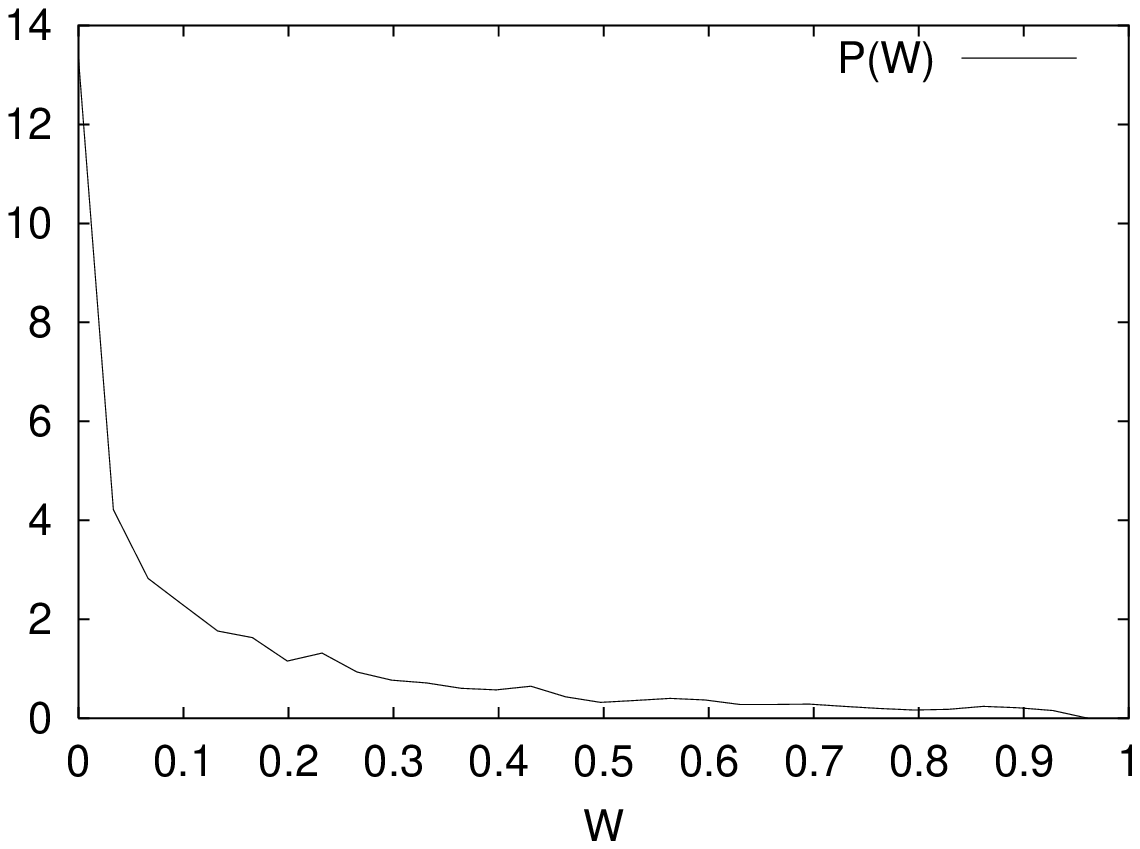}
\includegraphics[height=4cm,width=4.5cm]{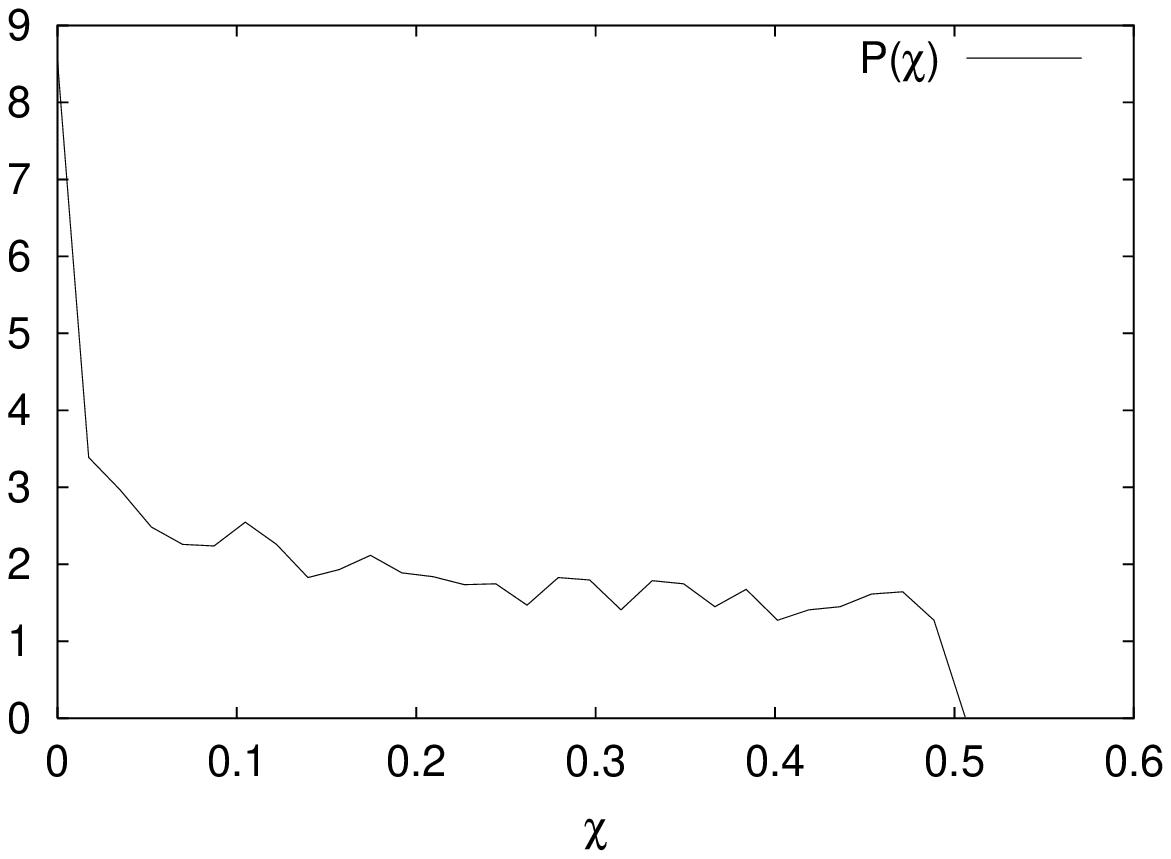}\\
\includegraphics[height=4cm,width=4.5cm]{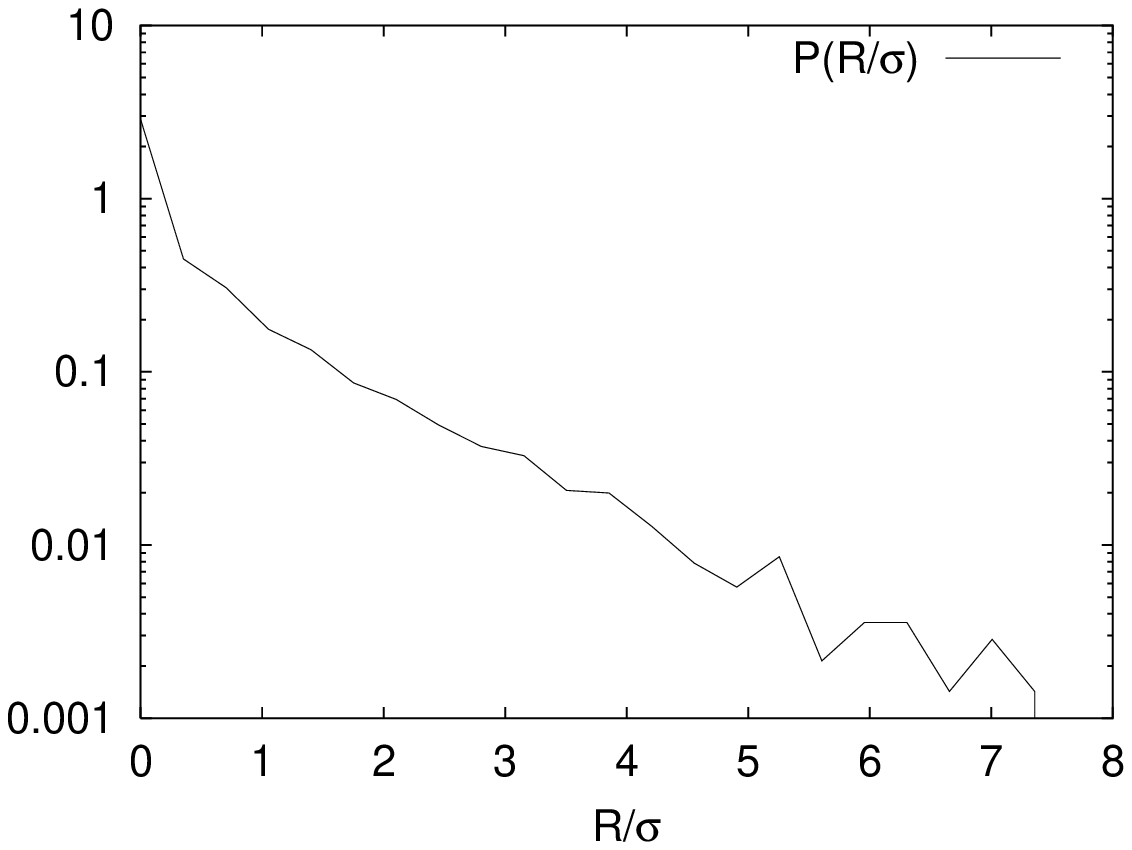}
\includegraphics[height=4cm,width=4.5cm]{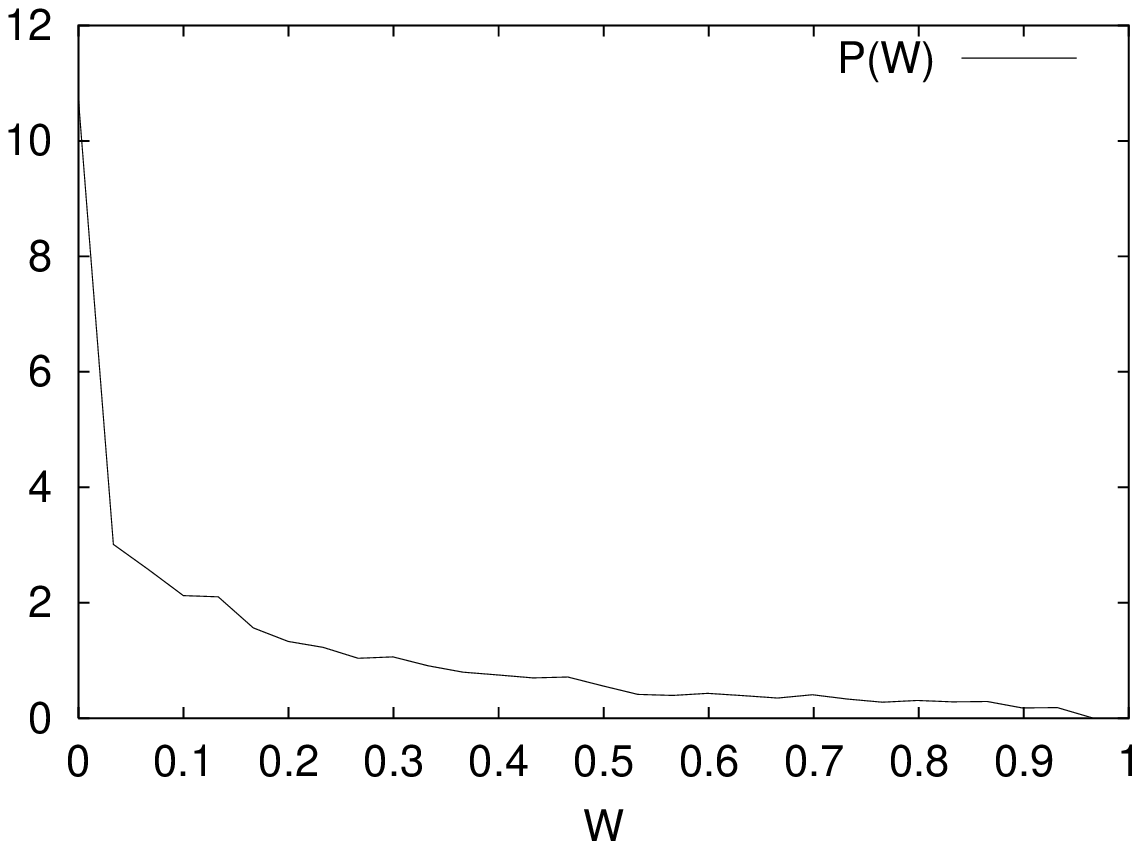}
\includegraphics[height=4cm,width=4.5cm]{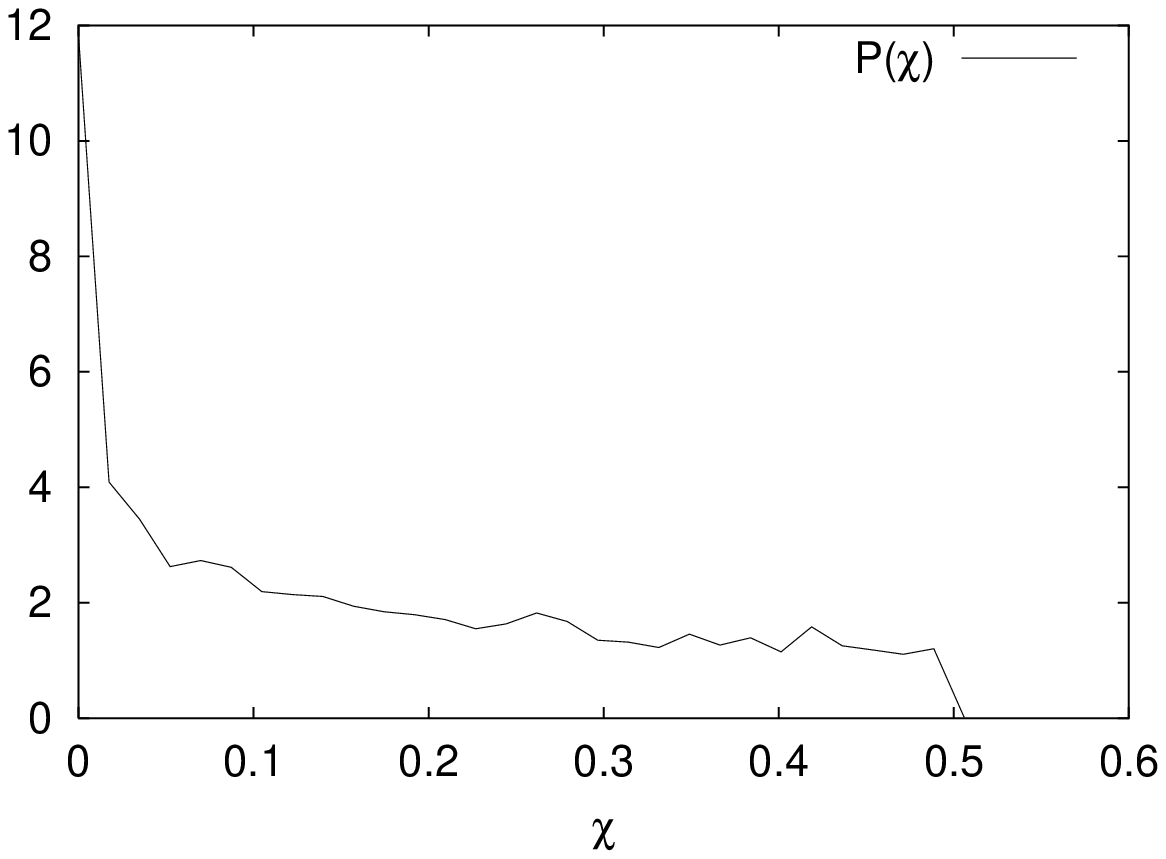}\\
\includegraphics[height=4cm,width=4.5cm]{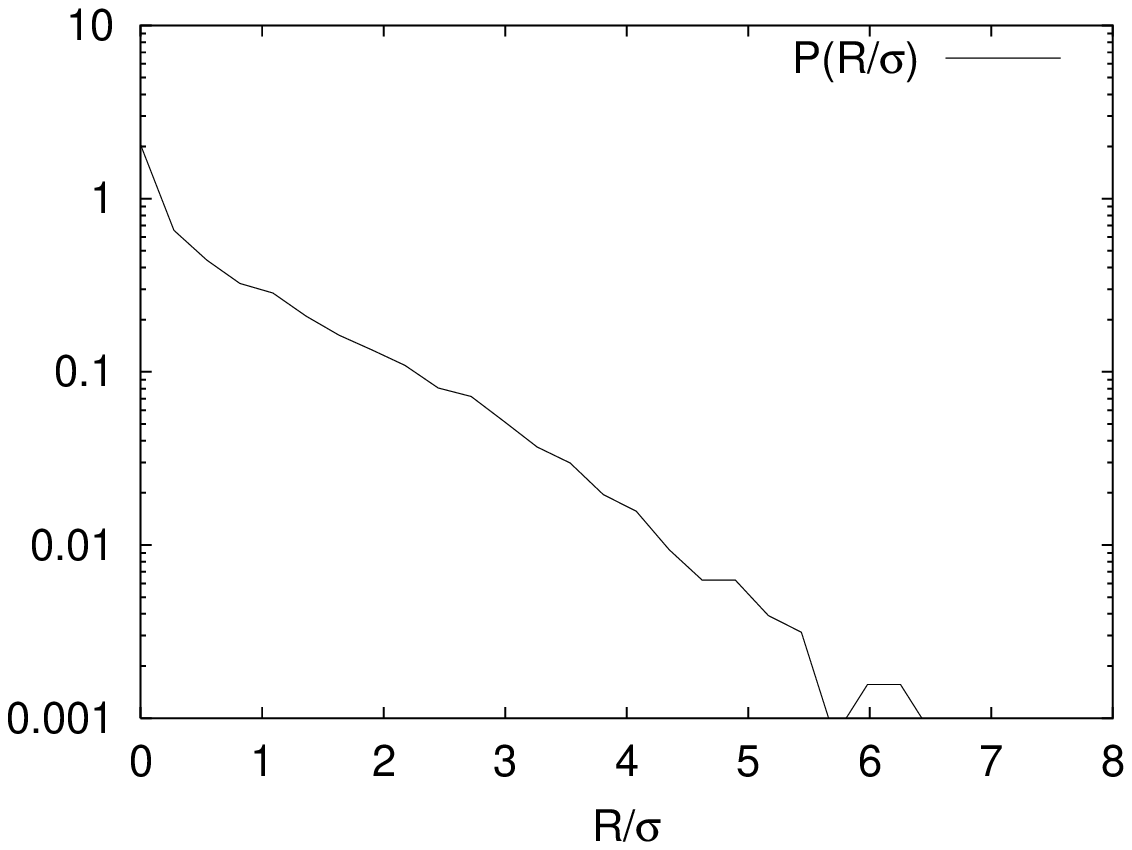}
\includegraphics[height=4cm,width=4.5cm]{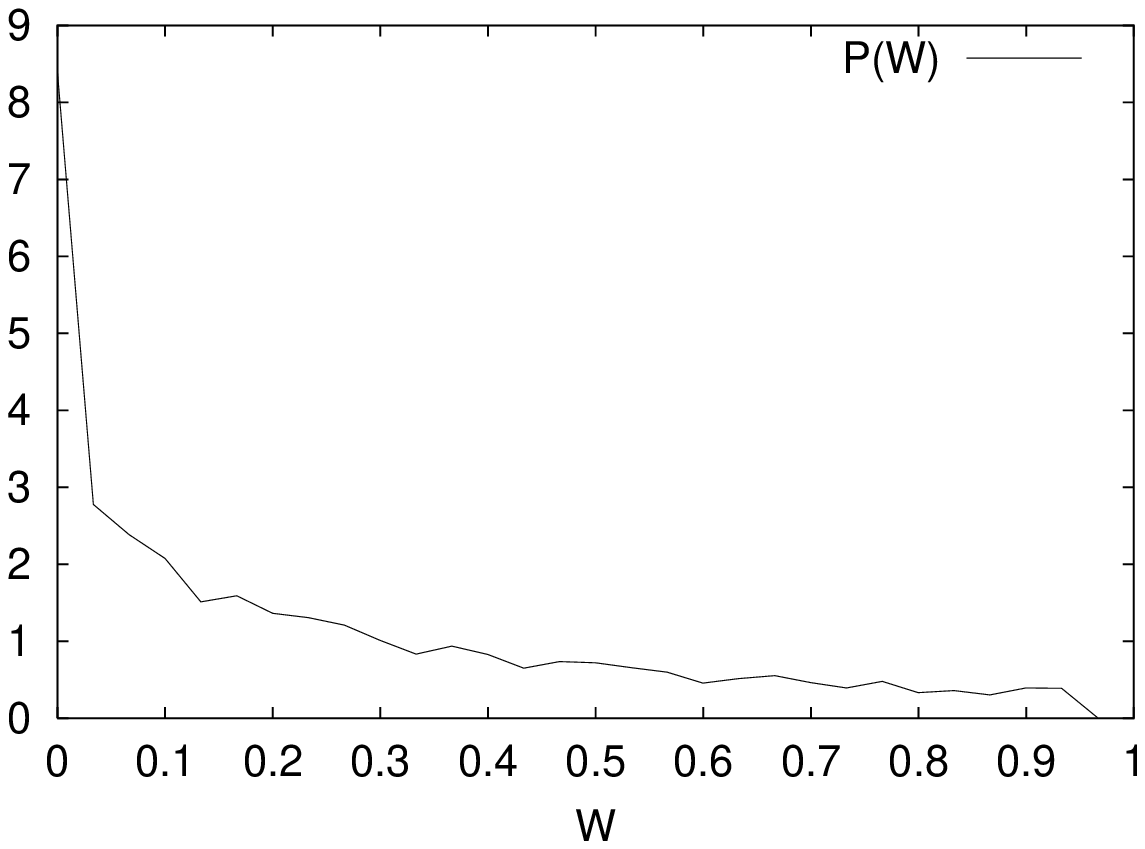}
\includegraphics[height=4cm,width=4.5cm]{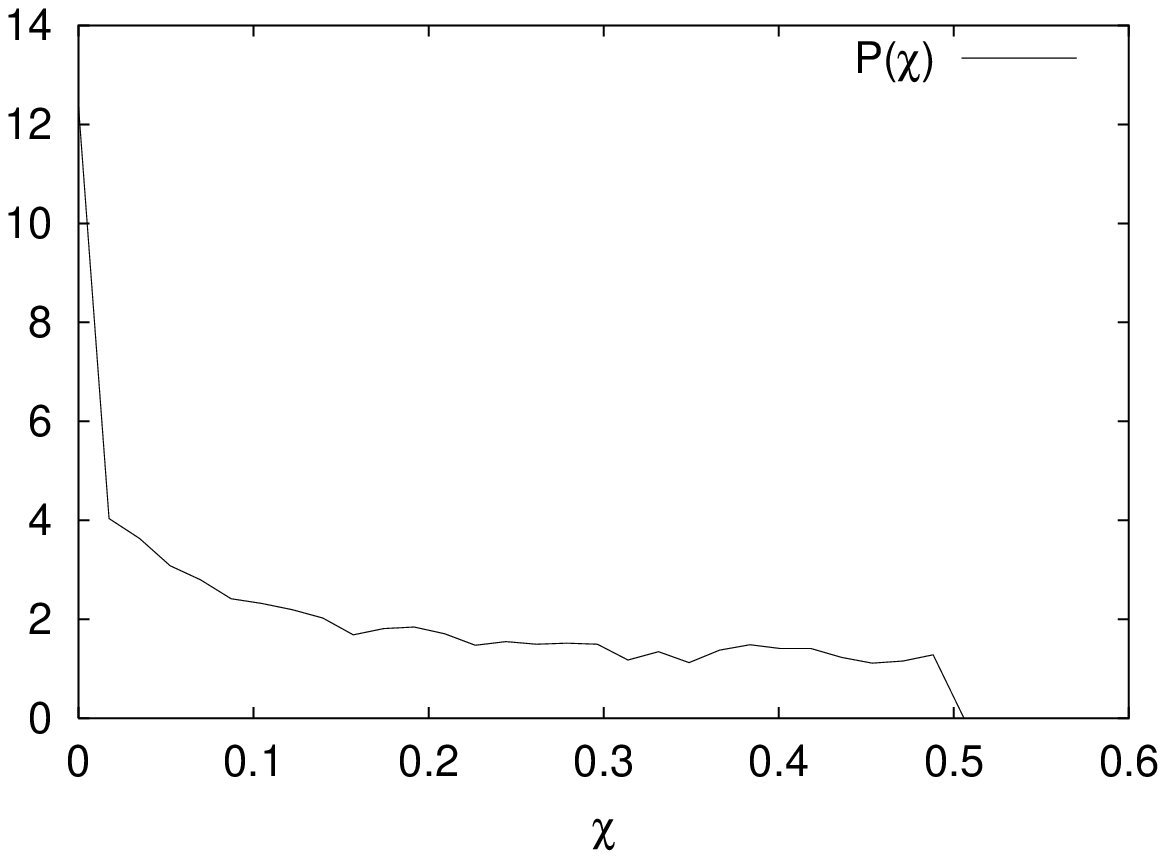}\\
\includegraphics[height=4cm,width=4.5cm]{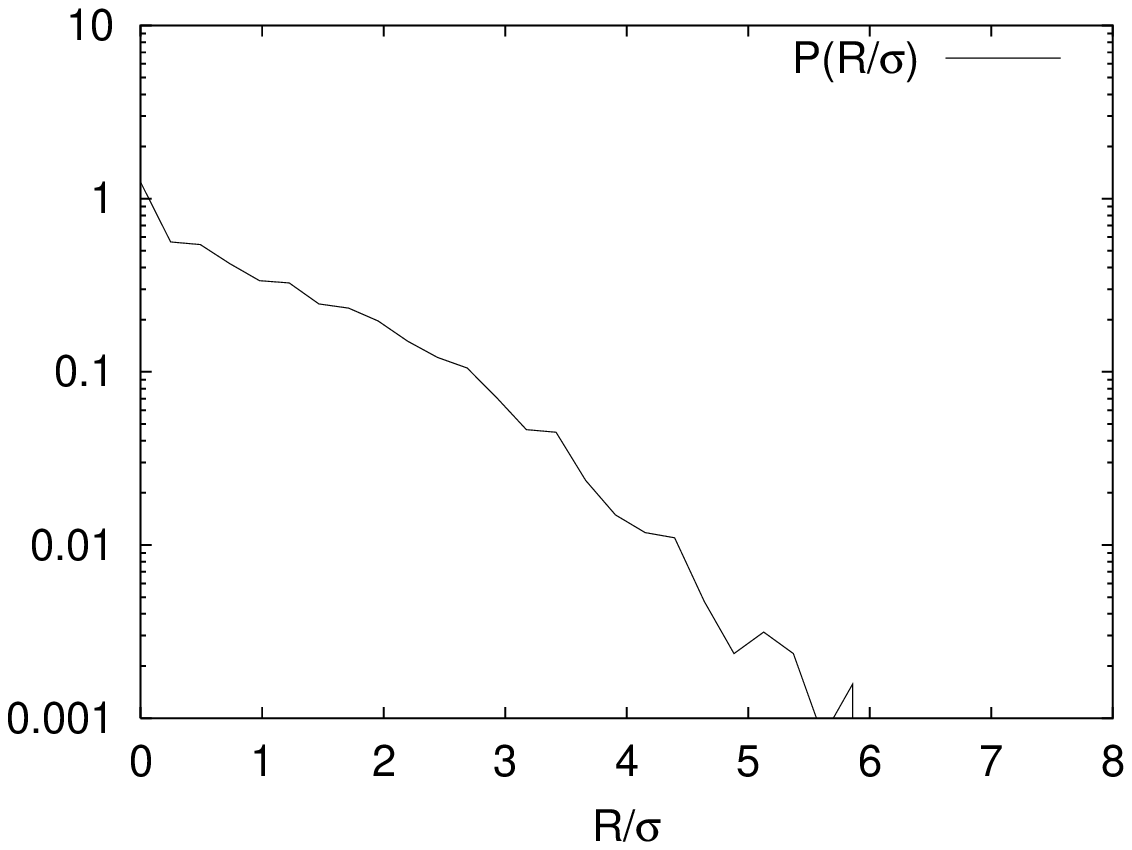}
\includegraphics[height=4cm,width=4.5cm]{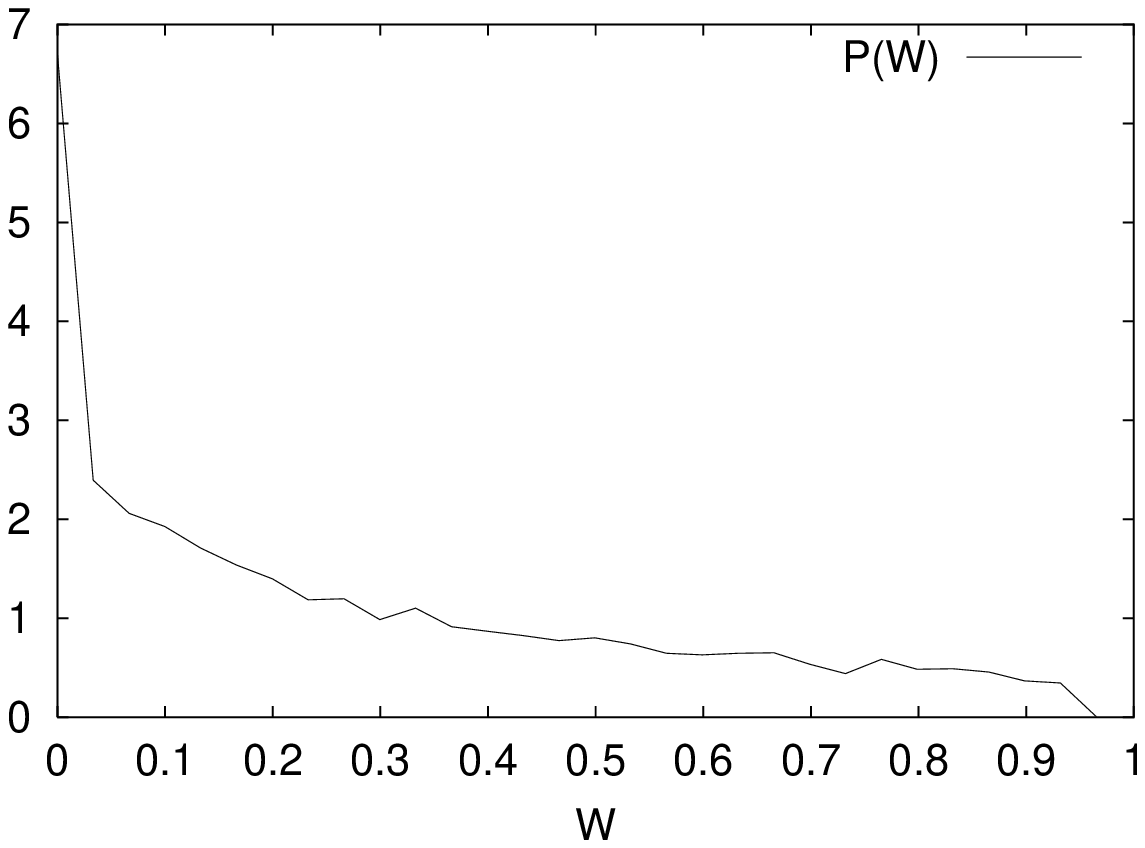}
\includegraphics[height=4cm,width=4.5cm]{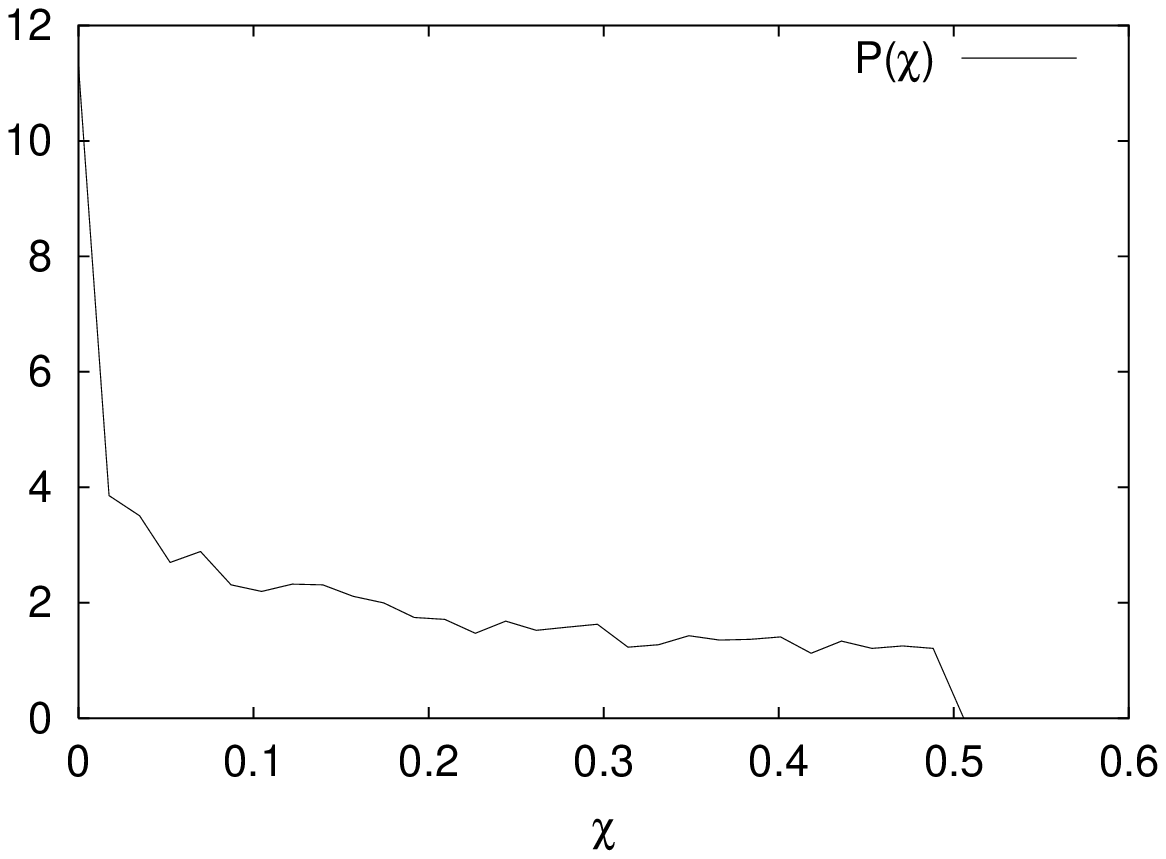}
\caption{\small{Time evolution of the PDF of $R/\sigma$, $w$ and
$\chi (rad)$ for $r_{0}/\eta=0.5$ in 2D KS, simulation with
$L/\eta=1691$ and $\lambda=0.5$ with $t_{E}/t_{\eta}=82.1$. From top to bottom the figures
are shown at times $t=20 \times t_{\eta}$, $t=60 \times t_{\eta}$,
$t=100 \times t_{\eta}$ and  $t=140 \times t_{\eta}$ respectively.}}
\label{Pdf_ks_1690}
\end{figure}
\begin{figure}
\includegraphics[height=5cm,width=7cm]{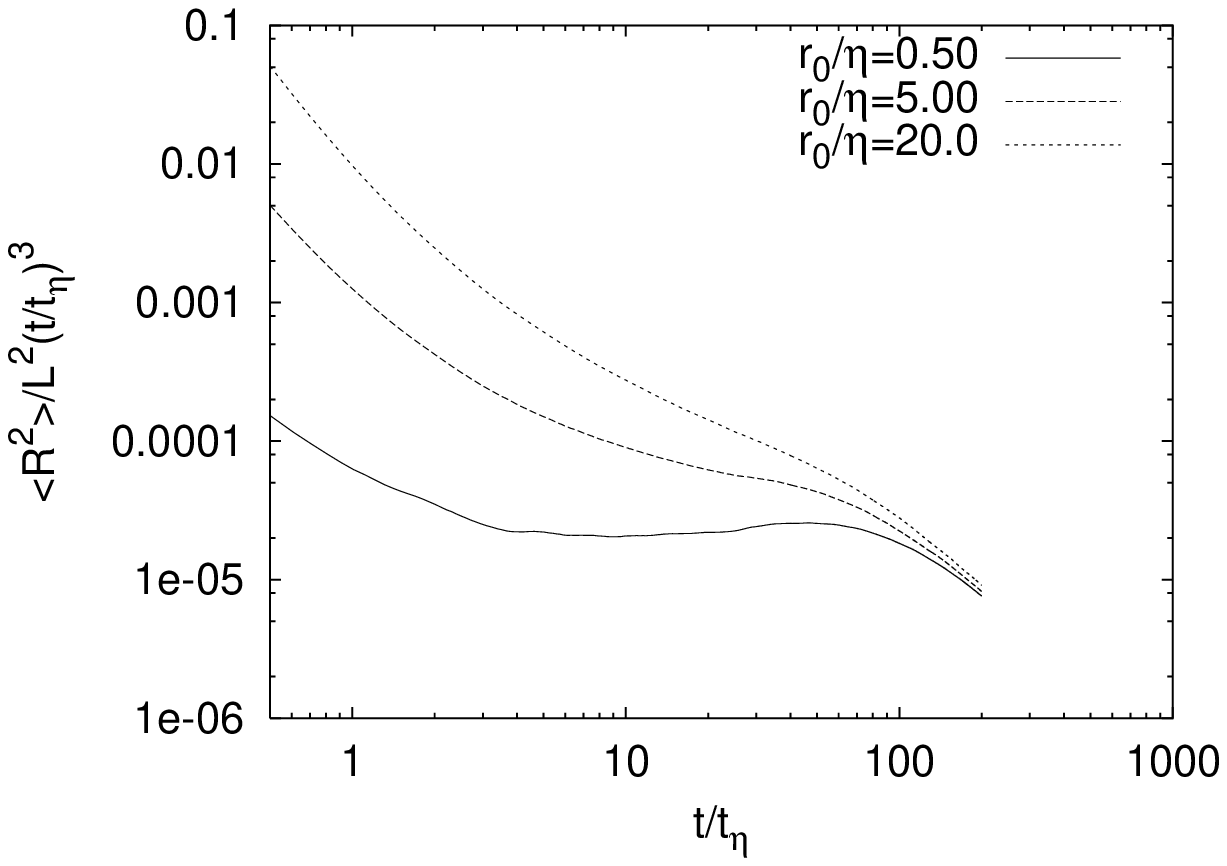}
\includegraphics[height=5cm,width=7cm]{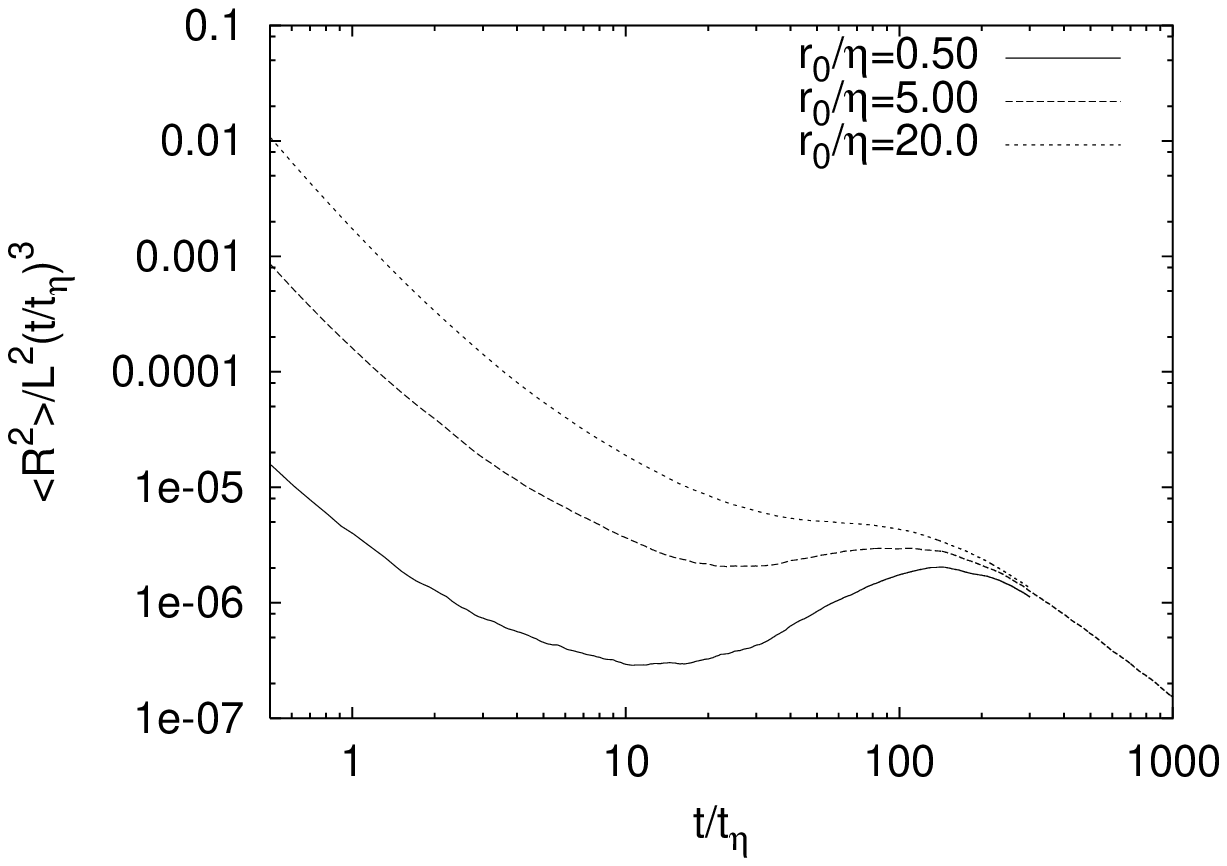}
\includegraphics[height=5cm,width=7cm]{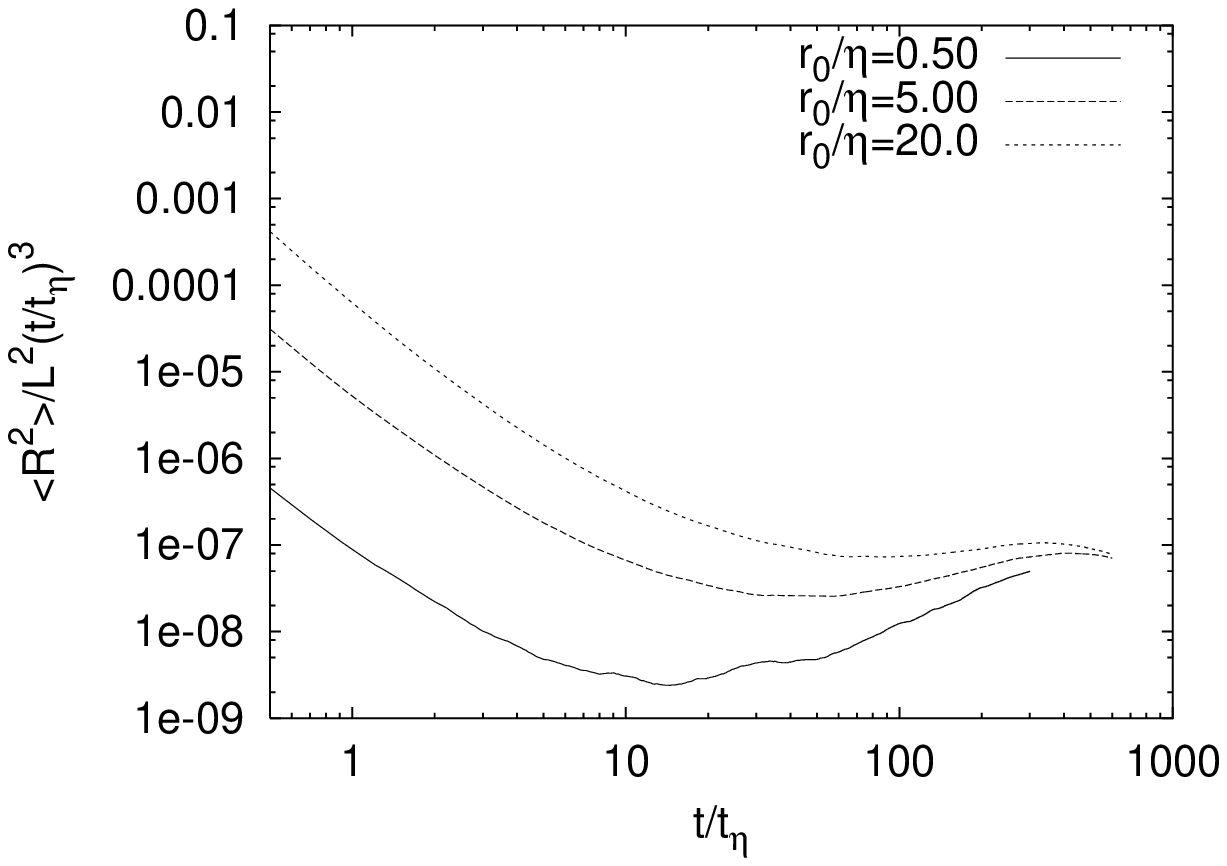}
\caption{\small{Time evolution of the $\langle
R(t)^{2}\rangle/(t/t_{\eta})^{3}$  obtained by Kinematic
Simulation of a triangular configuration of 3 particles in a two
dimensional high Reynolds number ($L/\eta=1691, 3381\ {\rm and}\
16909$) turbulent flow for different initial separation $r_0$'s
Here the energy spectrum $E(k)\sim k^{-5/3}$, $\lambda=0.5$, and
the number of realizations is $5\times 10^{3}$.}} \label{ks_tcube_1}
\end{figure}
\pagebreak
\begin{figure}
\includegraphics[height=5cm,width=7cm]{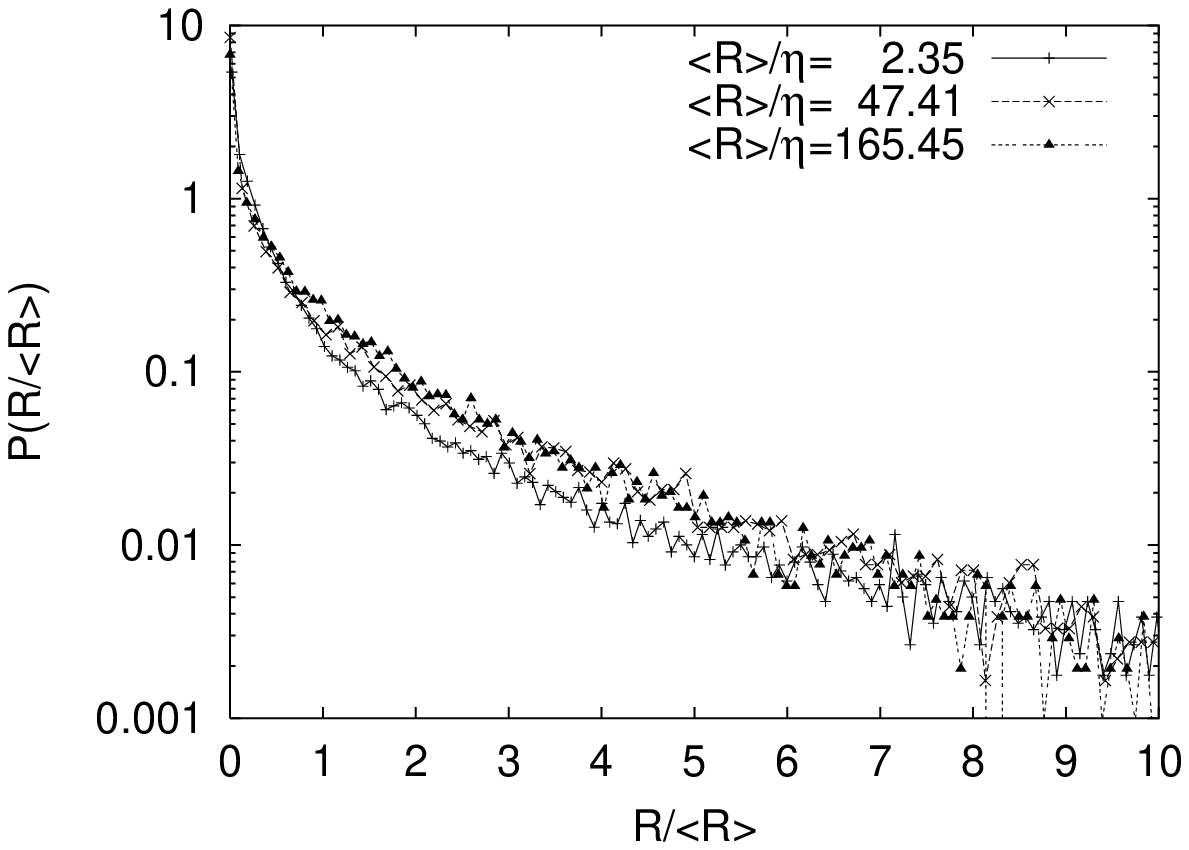}
\includegraphics[height=5cm,width=7cm]{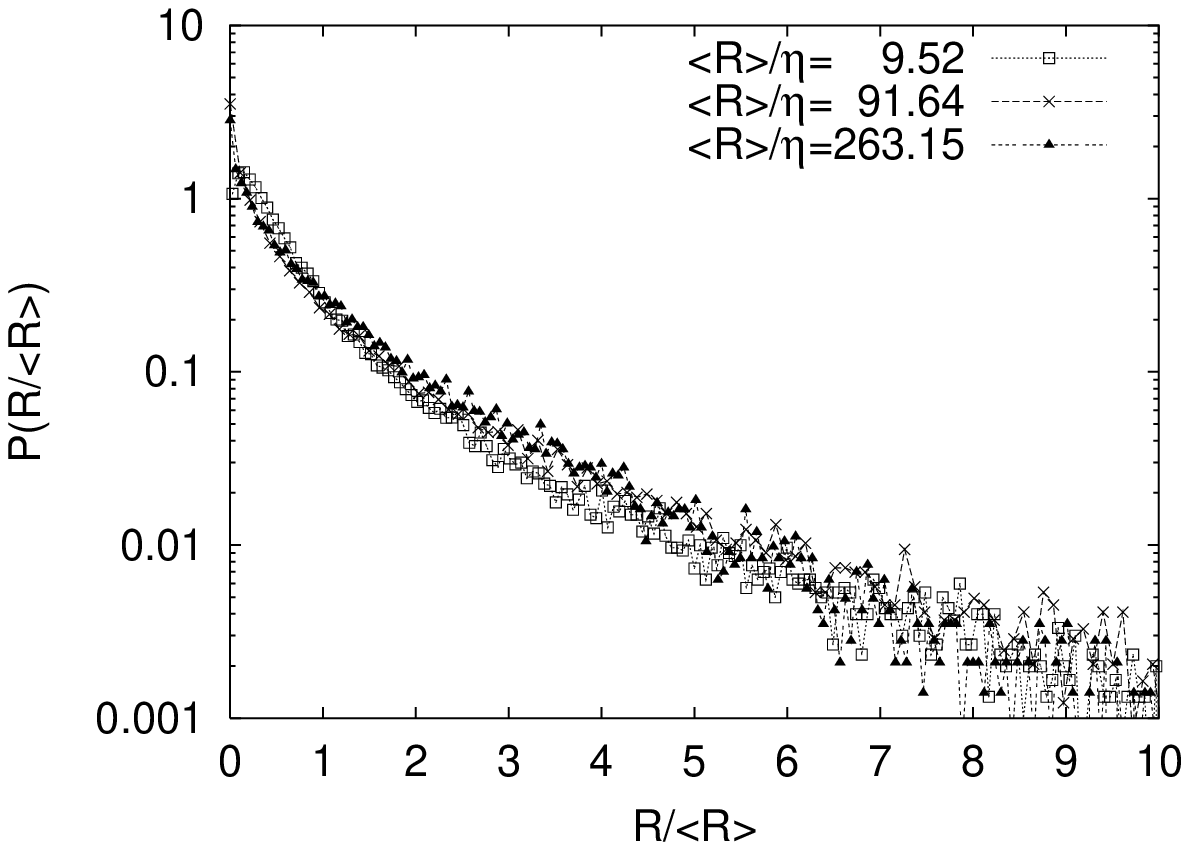}
\includegraphics[height=5cm,width=7cm]{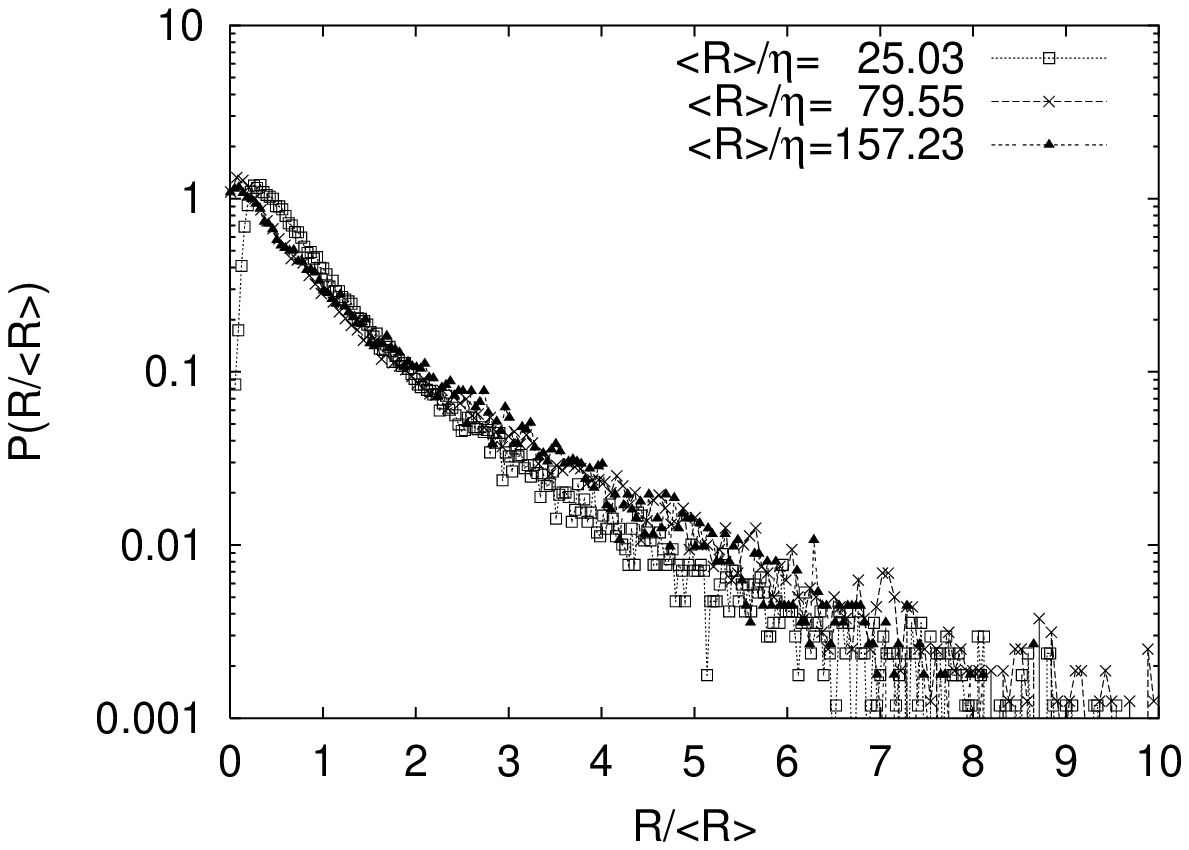}
\caption{\small{Time evolution of the PDF of $R$ with different
$r_{0}$'s obtained by KS in two
dimensions with $E(k)\sim k^{-5/3}$, $L/\eta=3381$. From left to right
$r_{0}/\eta=0.10, 1.0\ {\rm and}\ 5.0$.}}
\label{ks_pdf_evol}
\end{figure}
\begin{figure}
\includegraphics[height=5cm,width=7cm]{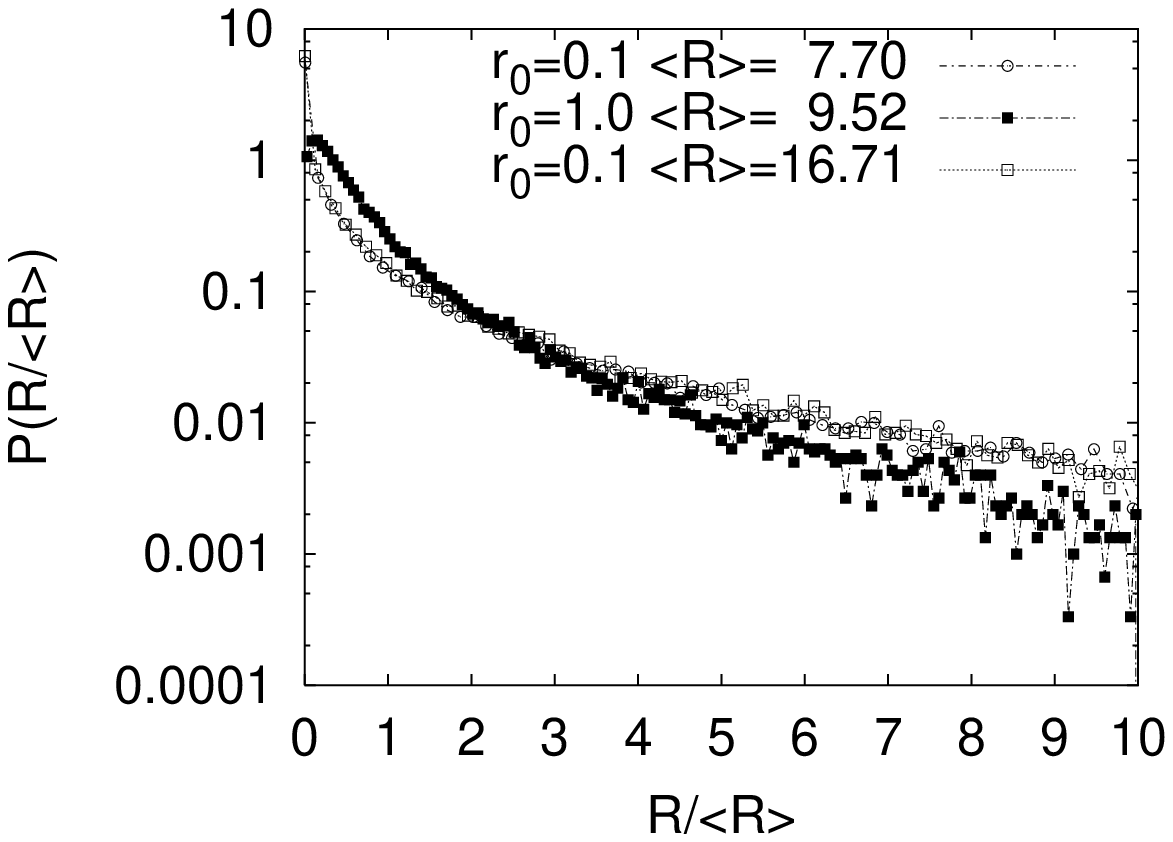}
\includegraphics[height=5cm,width=7cm]{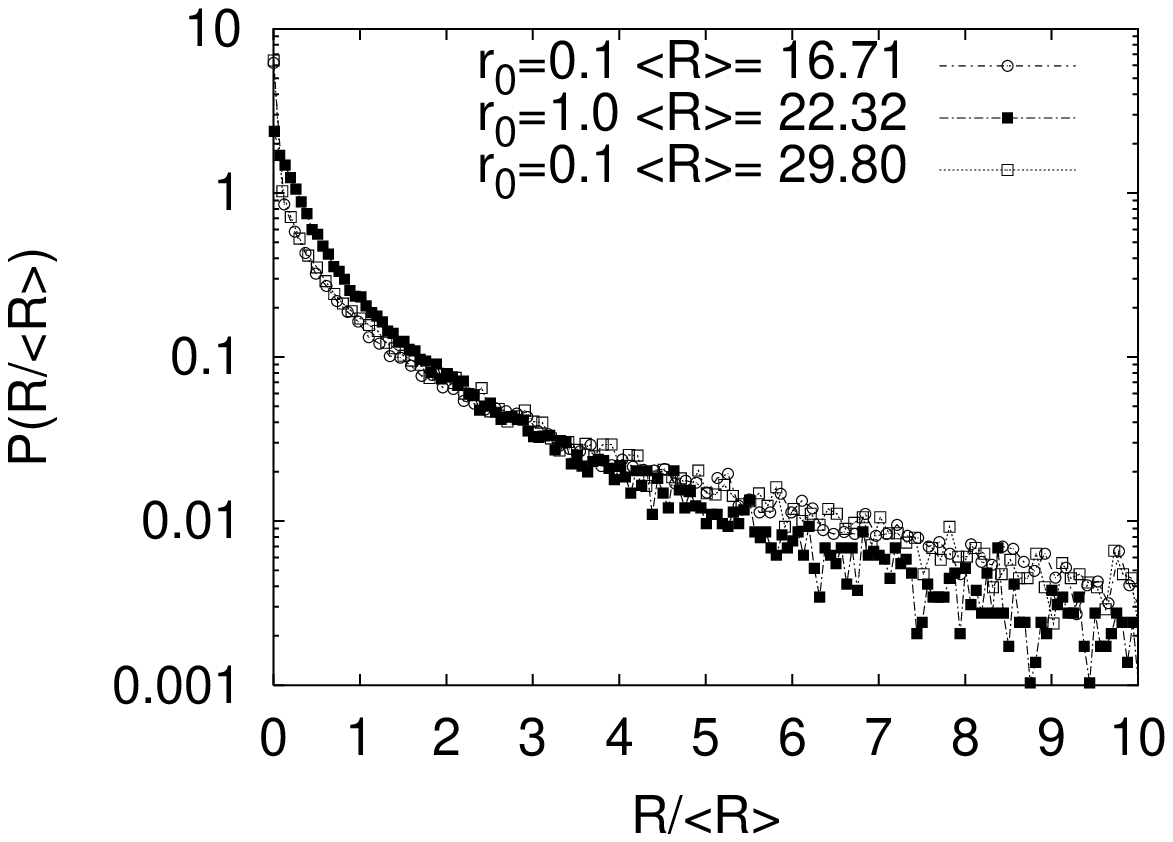}\\
\includegraphics[height=5cm,width=7cm]{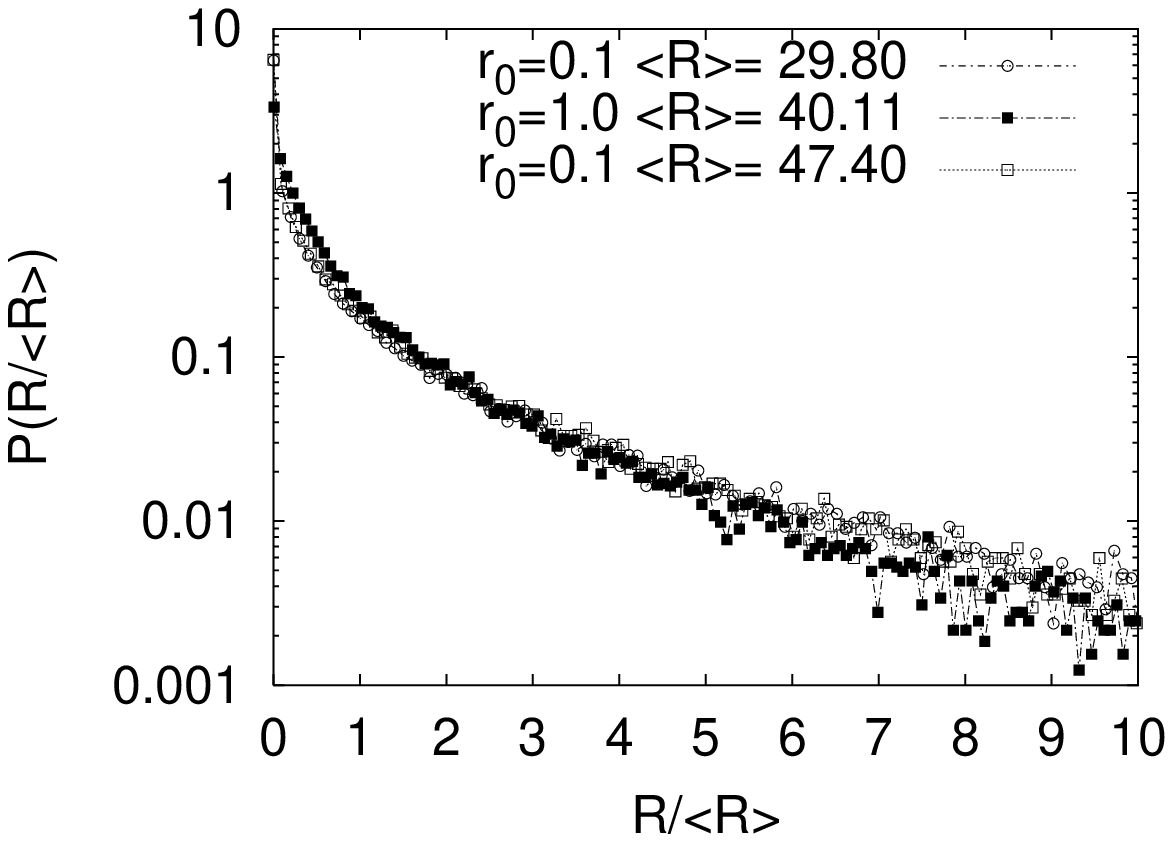}
\includegraphics[height=5cm,width=7cm]{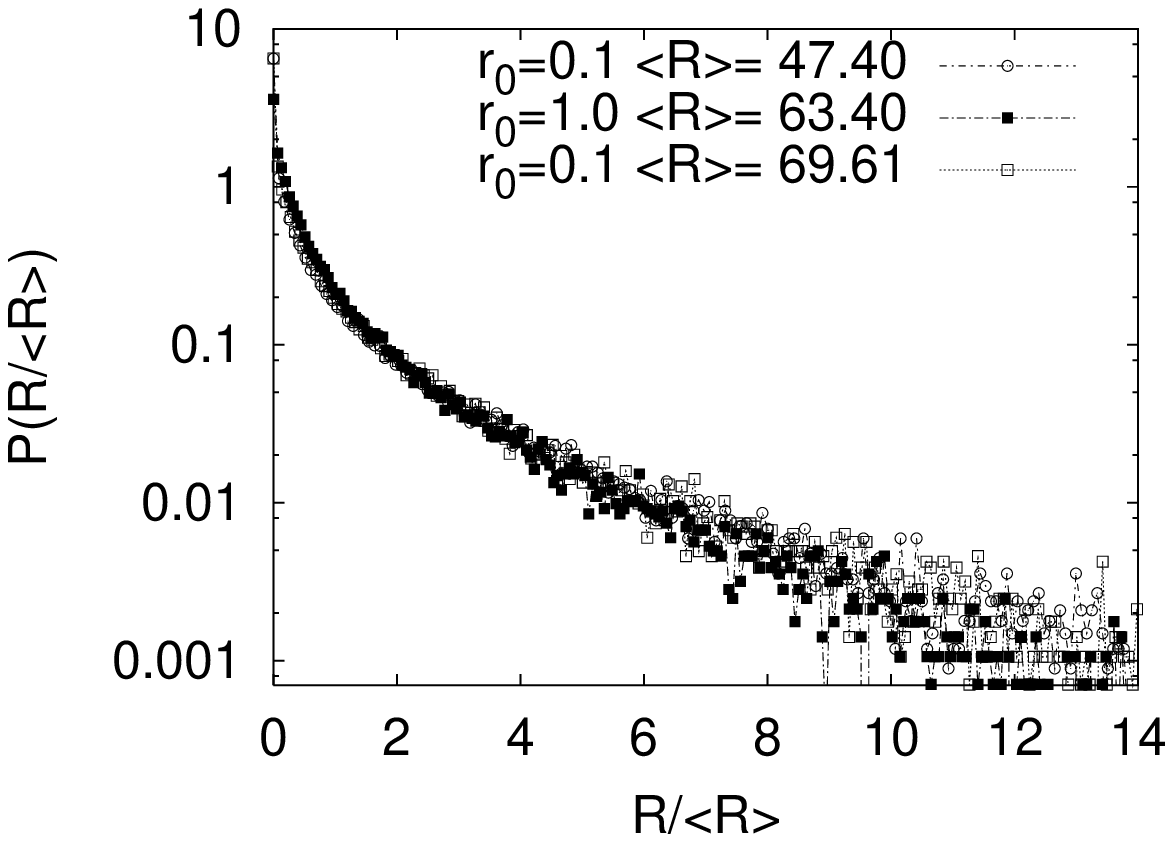}\\
\includegraphics[height=5cm,width=7cm]{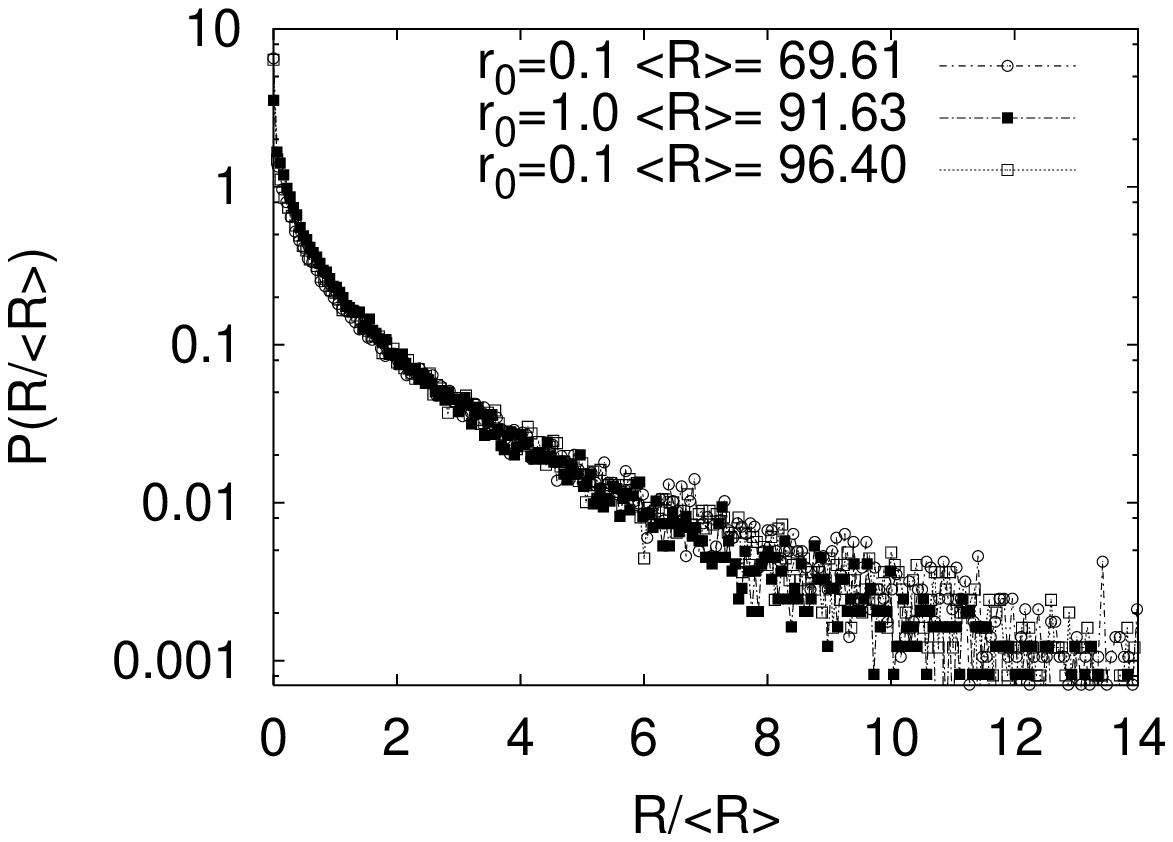}
\includegraphics[height=5cm,width=7cm]{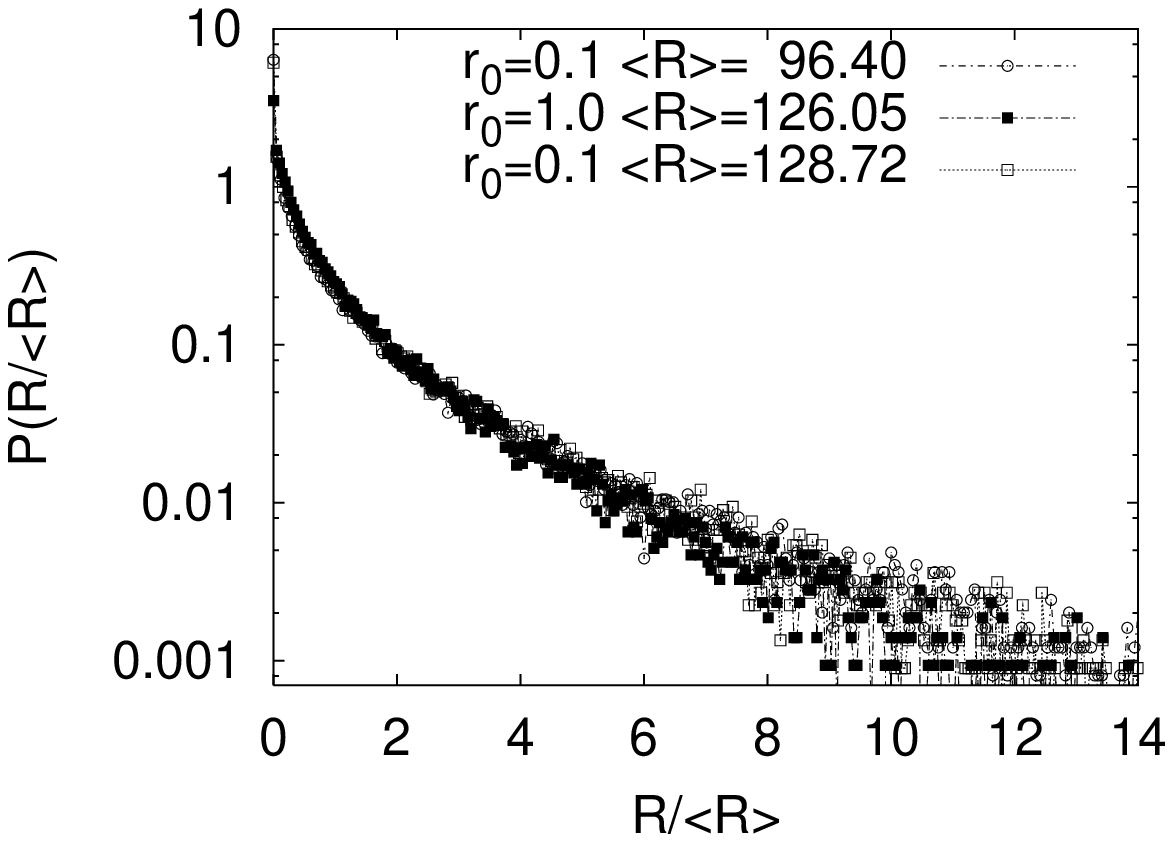}
\caption{\small{PDF of radius of gyration $R$ or global size with
different values of $r_{0}$. Here both $r_{0}$ and $\langle R\rangle$
have been normalized by $\eta$. Two runs with
different initial sizes are compared. The PDFs are shown at
several times, for $r_{0} = 0.1$ and $r_{0} = 1.0$.  For
comparison, the PDFs of $R$, corresponding to $r_0=0.1$ are
shown at two times where the value of $\langle R(t) \rangle$ are
the closest to the value of $\langle R(t) \rangle$ obtained with
the larger value  of $r_{0}$. The other parameters of the runs are
$L/\eta=3381$, $\lambda=0.5$ and $t_{E}/t_{\eta}=130.2$.}}
\label{ks_pdf_1}
\end{figure}
\begin{figure}
\includegraphics[height=5cm,width=7cm]{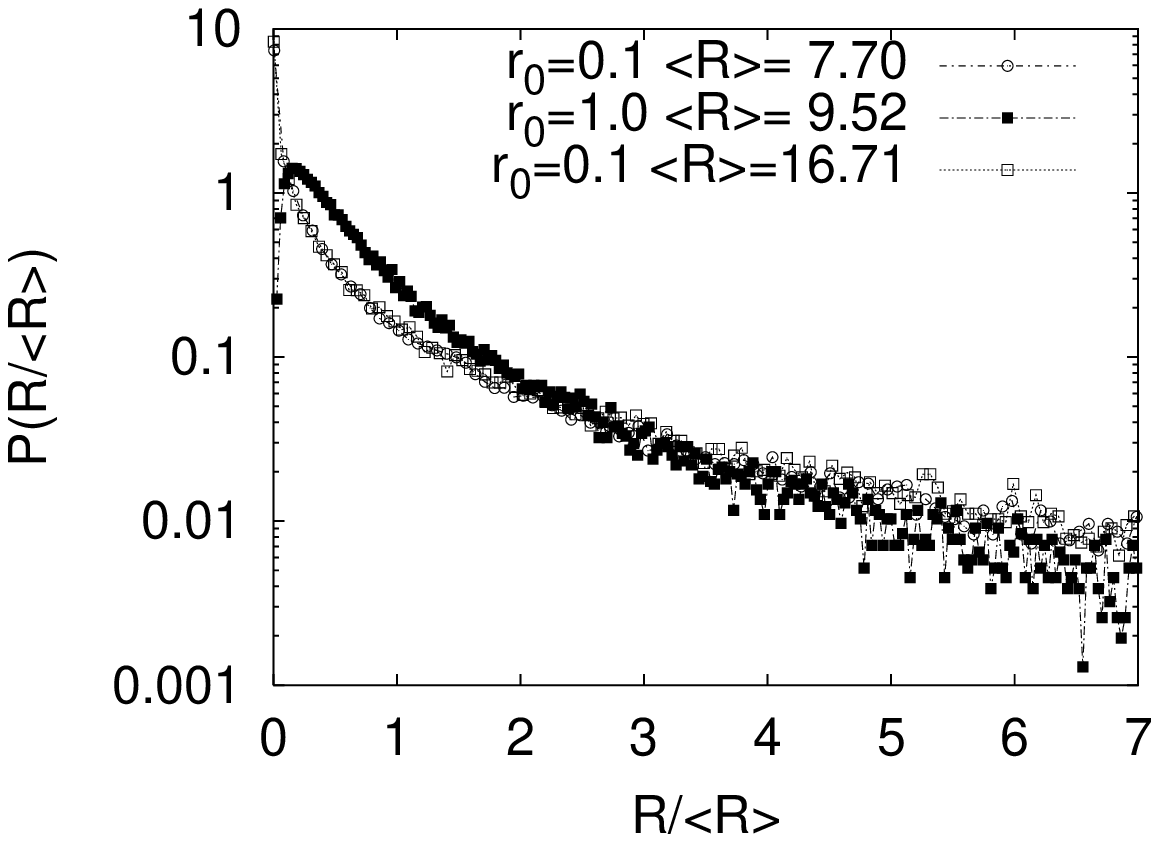}
\includegraphics[height=5cm,width=7cm]{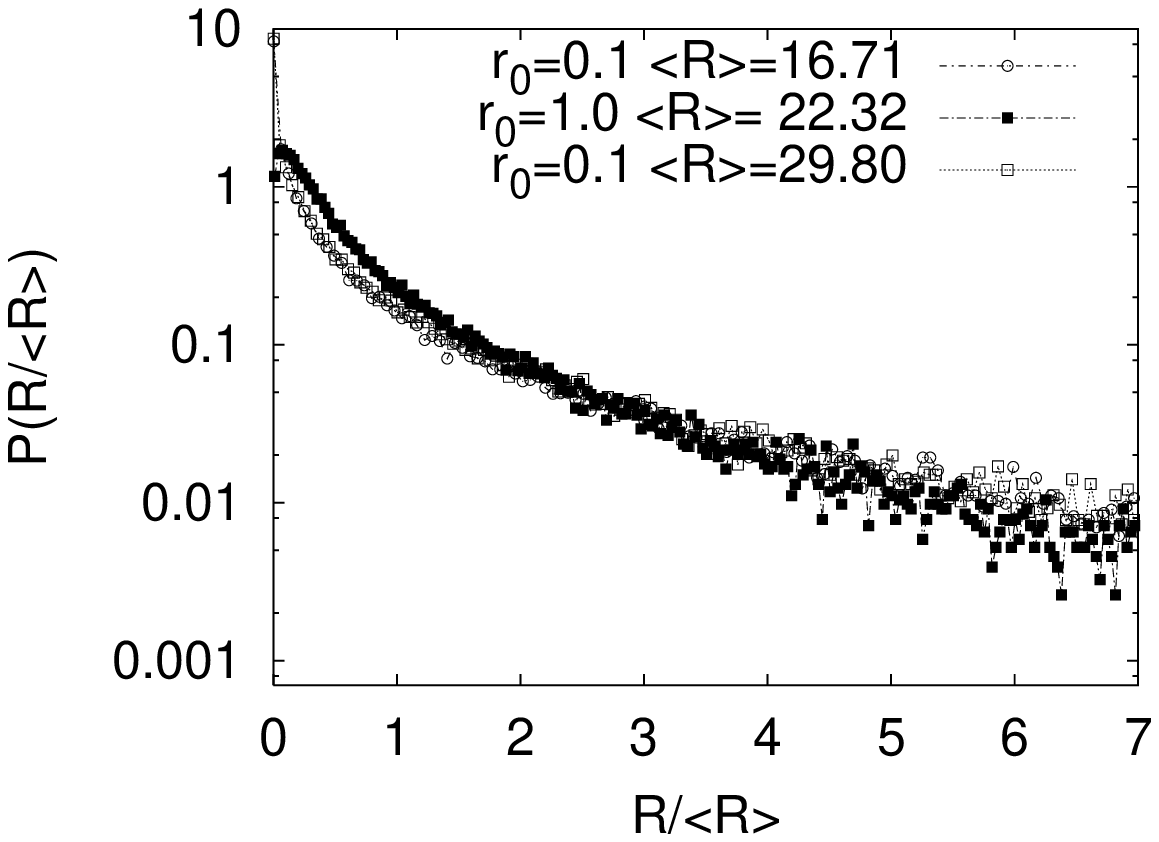}
\includegraphics[height=5cm,width=7cm]{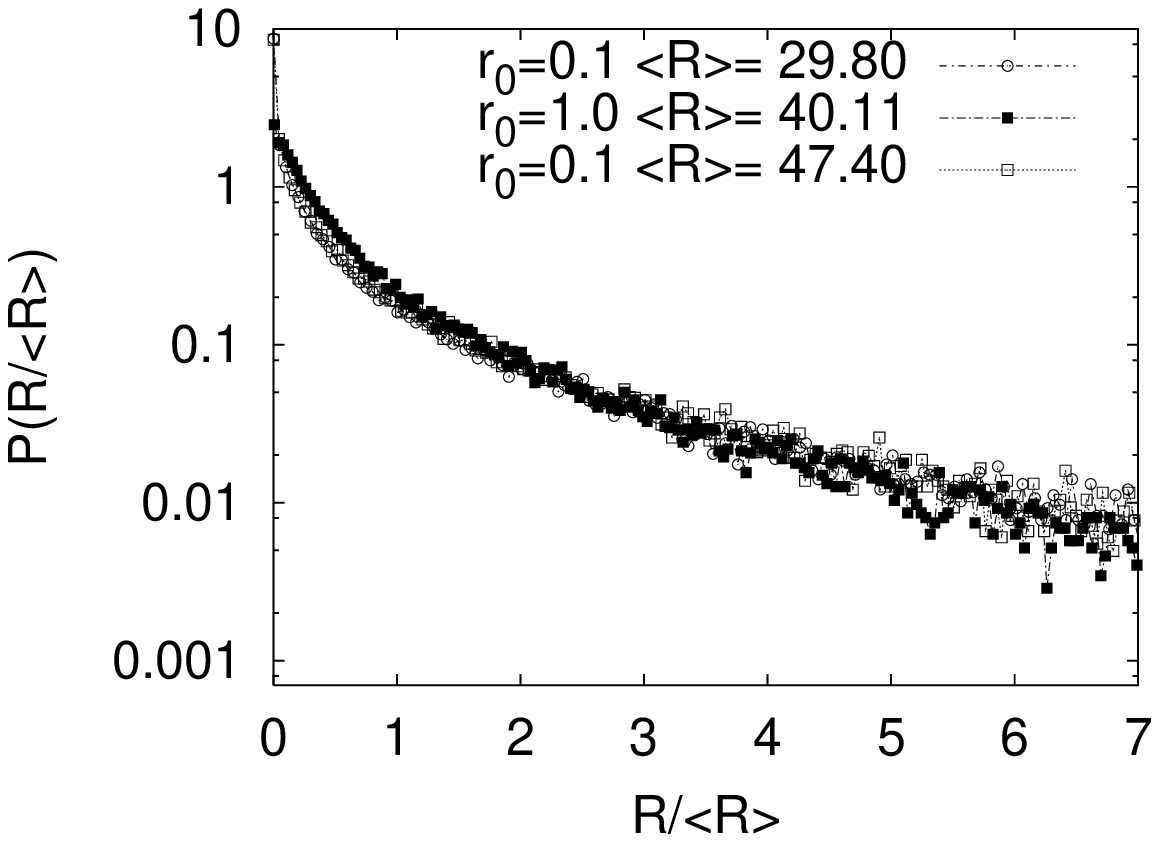}
\caption{\small{PDF of radius of gyration $R$ or global size with
different values of $r_{0}$. Here both $r_{0}$ and $\langle R\rangle$
have been normalized by $\eta$. Two runs with
different initial sizes are compared. The PDFs are shown at
several times, for $r_{0} = 0.1$ and $r_{0} = 1.0$.  For
comparison, the PDFs of $R$, corresponding to $r_0=0.1 $ are
shown at two times where the value of $\langle R(t) \rangle$ are
the closest to the value of $\langle R(t) \rangle$ obtained with
the larger value  of $r_{0}$. The other parameters of the runs are
$L/\eta=3381$, $\lambda=0.5$ and $t_{E}/t_{\eta}=130.2 $.
The plots in this figure are the zoomed in version of figure
\ref{ks_pdf_1}.}} \label{ks_pdf_2}
\end{figure}
\begin{figure}
\includegraphics[height=5cm,width=7cm]{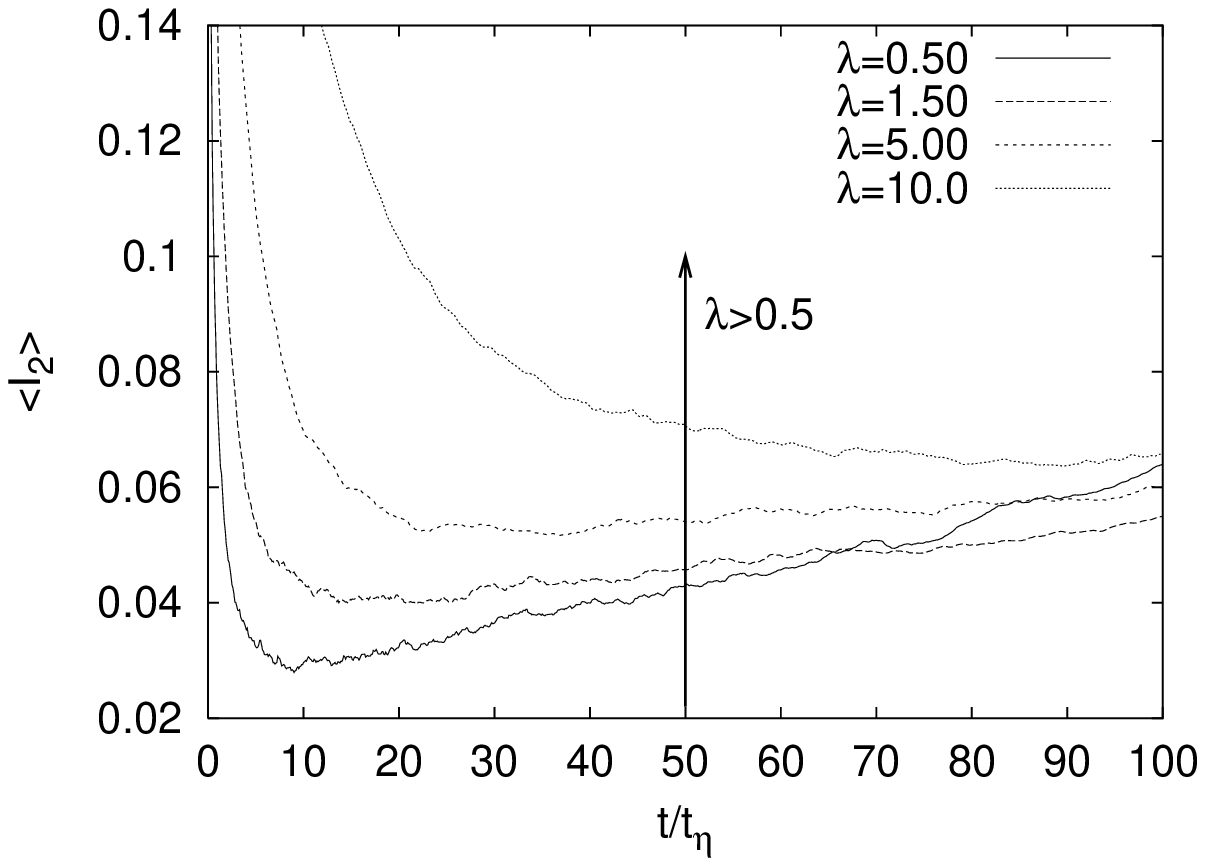}
\includegraphics[height=5cm,width=7cm]{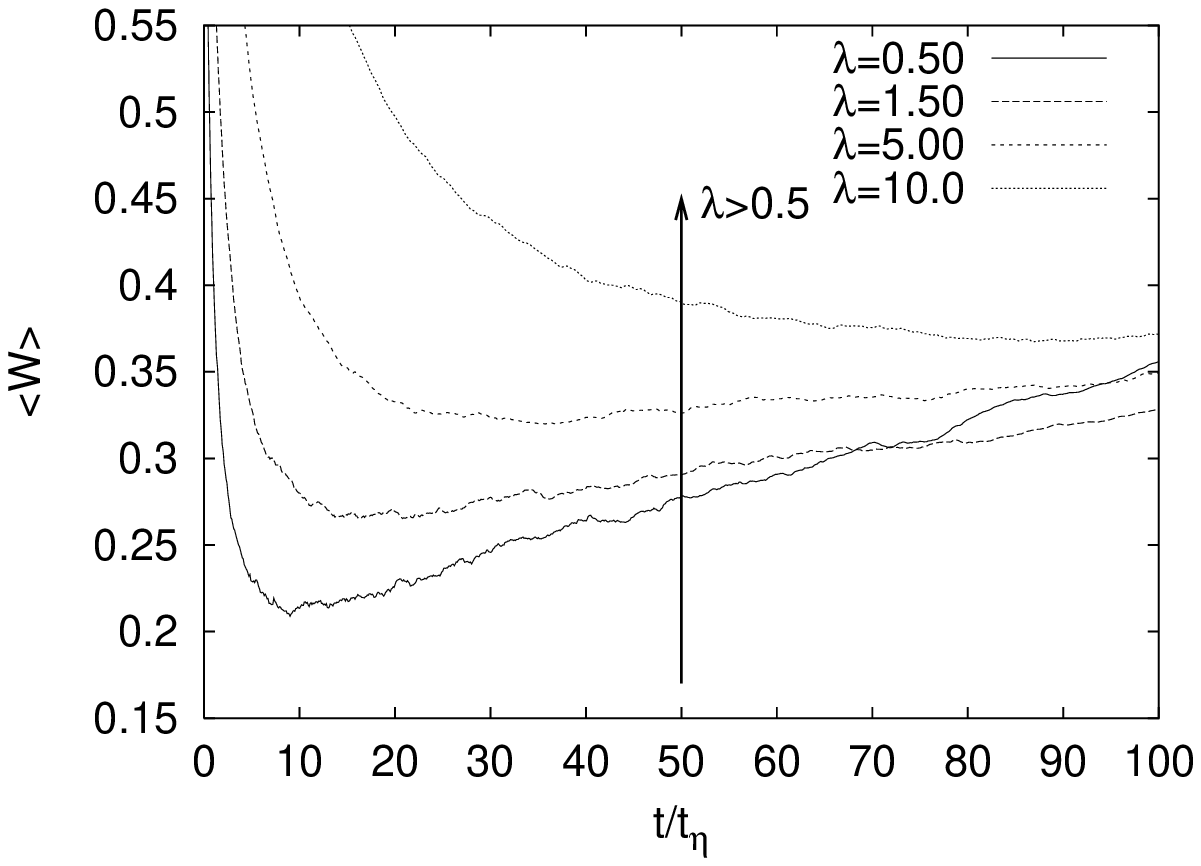}
\caption{\small{Time evolution of $\langle I_{2}(t)\rangle$ and
$\langle w(t)\rangle$ obtained by Kinematic Simulation of a triangular
configuration of 3 particles in a two dimensional high Reynolds
number ($L/\eta=1691$) turbulent flow for different
$\lambda$'s. Here the energy spectrum $E(k)\sim k^{-5/3}$, initial
separation $r_0=0.5\times\eta$ and number of realizations is
$5\times10^{3}$.}}
\label{I2W_lambda}
\end{figure}
\begin{figure}
\includegraphics[height=5cm,width=7cm]{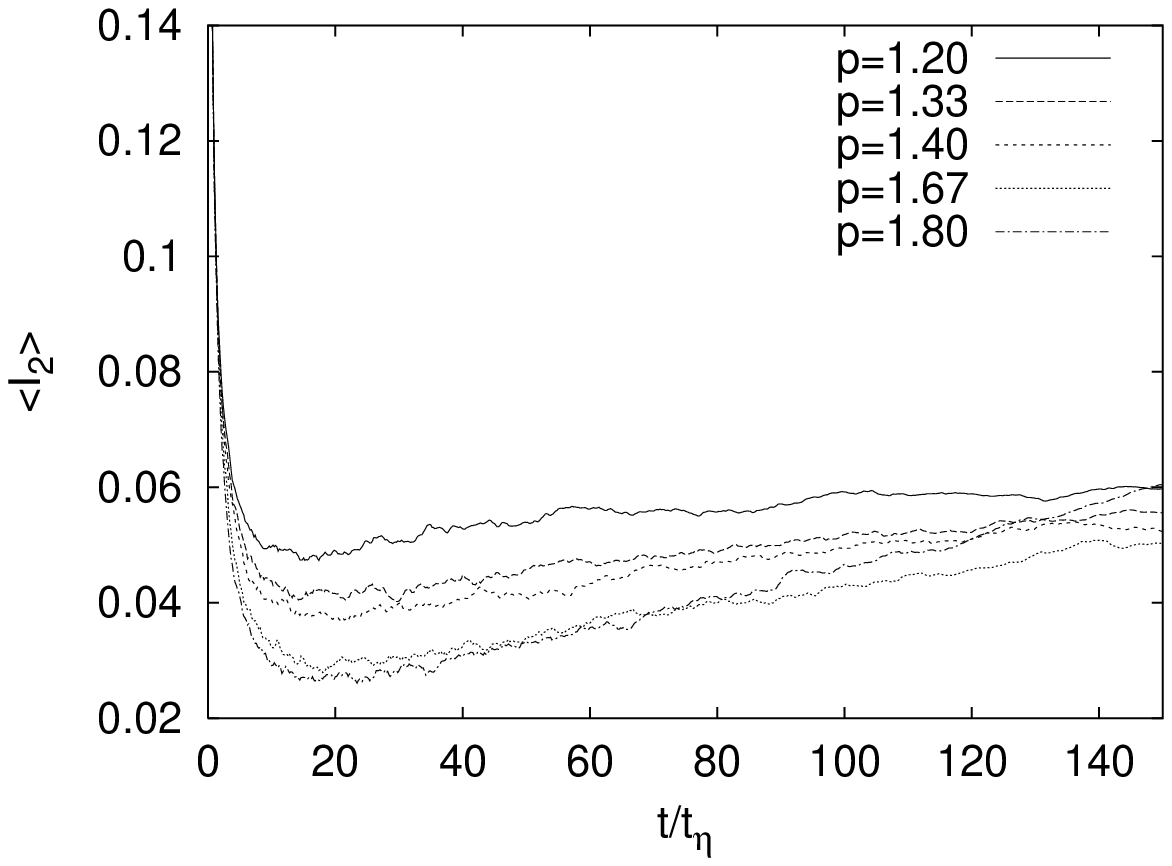}
\includegraphics[height=5cm,width=7cm]{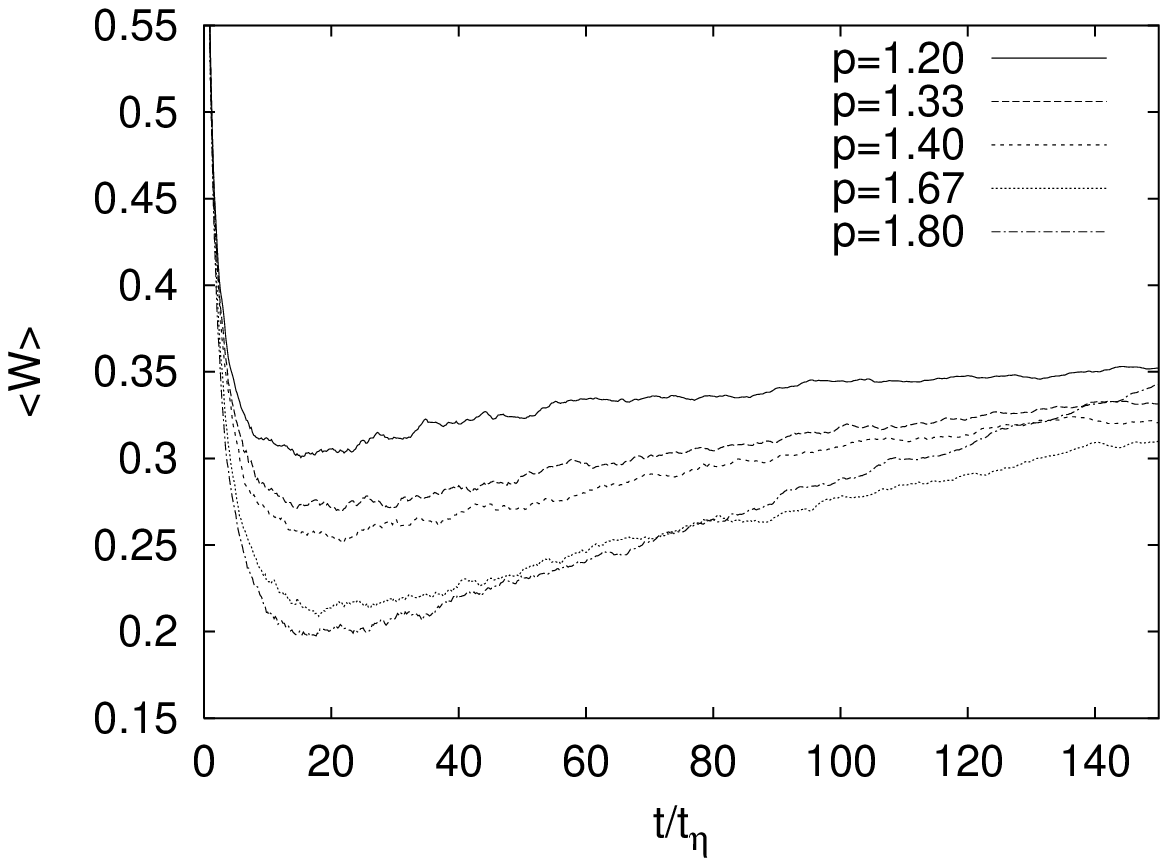}
\caption{\small{Time evolution of $\langle I_{2}(t)\rangle$ and
$\langle w(t)\rangle$ obtained by Kinematic Simulation of a triangular
configuration of 3 particles in a two dimensional high Reynolds number
($L/\eta=1691$) turbulent flow for different energy
spectra $E(k)\sim k^{-p}$. Here $\lambda=0.5$, initial separation
$r_0=0.5\ \eta$ and number of realizations is $5\times10^{3}$.}}
\label{I2W_p}
\end{figure}

\begin{table}
\begin{center}
\begin{tabular}{|c|c|c|c|c|c|c|}
\hline
No. of&$L/\eta$ &No. of& $r_{0}/\eta$&$\lambda$&No. of&$E(k)\sim k^{-p}$ \\
Runs& &Triangles& & &Modes $N_{k}$& $p$\\
\hline
\hline
$1$ &$3.67$ &$1\times 10^{4}$ &$1/6, 2/3, 8/3, 4$&$0.5$&$79$& $1.67$\\
$2$ &$10$   &$1\times 10^{4}$ &$0.05, 0.5, 5$&$0.5$&$79$& $1.67$\\
$3$ &$1691$ &$1\times 10^{4}$ &$0.5, 5,20,64$&$0.5$&$79$&$1.67$\\
$4$ &$1691$ &$1\times 10^{4}$ &$0.5$ &$1.5$&$79$&$1.67$\\
$5$ &$1691$ &$1\times 10^{4}$ &$0.5$ &$5.0$&$79$&$1.67$\\
$6$ &$1691$ &$1\times 10^{4}$ &$0.5$ &$10.0$&$79$&$1.67$\\
$7$ &$1691$ &$1\times 10^{4}$ &$0.5$ &$0.5$&$79$&$1.20$\\
$8$ &$1691$ &$1\times 10^{4}$ &$0.5$ &$0.5$&$79$&$1.33$\\
$9$ &$1691$ &$1\times 10^{4}$ &$0.5$ &$0.5$&$79$&$1.40$\\
$10$&$1691$ &$1\times 10^{4}$ &$0.5$ &$0.5$&$79$&$1.80$\\
$11$&$3380$ &$5\times 10^{4}$ &$5, 20, 64$ &$0.5$&$200$&$1.67$\\
$12$&$11180$&$1\times 10^{4}$ &$5, 64$ & $0.5$&$500$&$1.67$\\
$13$&$16909$&$1\times 10^{4}$ &$5,20, 64$ & $0.5$&$500$&$1.67$\\
\hline
\end{tabular}
\caption{\label{table_3_part}Different simulation parameters of our runs with kinematic
simulation.}
\end{center}
\end{table}

\end{document}